\documentclass{emulateapj}

\usepackage{lscape}
\usepackage{natbib}
\def\S2.3{$S_{2.3}$}

\shorttitle{Star Forming Galaxy Templates}
\shortauthors{G. H. Rieke et al.}

\begin{document}

\title{Determining Star Formation Rates for Infrared Galaxies}

\author{G. H. Rieke \altaffilmark{1},
A. Alonso-Herrero \altaffilmark{2},
B. J. Weiner \altaffilmark{1},
P. G. P\'erez-Gonz\'alez \altaffilmark{1,3},
M. Blaylock \altaffilmark{4},
J. L. Donley \altaffilmark{1},
D. Marcillac \altaffilmark{5}}

\altaffiltext{1}{Steward Observatory, University of Arizona, 933 North Cherry Avenue, 
Tucson, AZ 85721, USA}
\altaffiltext{2}{Dpto. Astrofisica Molecular e Infrarroja, Instituto de Estructura 
de la Materia, CSIC, Madrid, Spain}
\altaffiltext{3}{Universidad Complutense de Madrid, Facultad de Ciencias F\'isicas
Dpto. de Astrofísica y CC. de la Atm\'osfera, Madrid 28040 Spain}
\altaffiltext{4}{Department of Mechanical and Aero Engineering, Univ. of Cal., Davis, CA 95616, USA}
\altaffiltext{5}{Institut d'Astrophysique Spatiale (IAS), b\^atiment 121, Universit\'e Paris-Sud 11, France}

\begin{abstract}

We show that measures of star formation rates (SFRs) for infrared galaxies
using either single-band 24$\mu$m or extinction-corrected Pa$\alpha$ luminosities
are consistent in the total infrared luminosity = L(TIR) $\sim$ 10$^{10}$ L$_\odot$ range. 
MIPS 24$\mu$m photometry can yield star formation rates
accurately from this luminosity upward: SFR(M$_\odot$ yr$^{-1}$)=$7.8 \times 10^{-10}$ L(24$\mu$m, L$_\odot$)
from L(TIR) = $5 \times$ 10$^9$ L$_\odot$ to 10$^{11}$ L$_\odot$ and 
SFR = $7.8 \times 10^{-10}$ L(24$\mu$m, L$_\odot$)($7.76 \times 10^{-11}$ L(24))$^{0.048}$ 
for higher L(TIR). For galaxies with L(TIR) $\ge$ 10$^{10}$ L$_\odot$, 
these new expressions should provide SFRs to within 0.2 dex. 
For L(TIR) $\ge$ 10$^{11}$ L$_\odot$, we find that the SFR of infrared galaxies is significantly
underestimated using extinction-corrected Pa$\alpha$ (and presumably using any
other optical or near infrared recombination lines). 
As a part of this work, we constructed spectral energy distribution (SED) templates for eleven luminous and 
ultraluminous purely star forming infrared galaxies (LIRGs and ULIRGs) and over the spectral range
0.4$\mu$m to 30 cm. We use these templates and the SINGS data 
to construct average templates from 5$\mu$m to 30 cm for infrared galaxies
with L(TIR) = $5 ~\times$ 10$^{9}$ to 10$^{13}$ L$_\odot$. 
All of these templates are made available on line.

\end{abstract}

\section{Introduction}

The rate at which a galaxy is forming massive stars is a central measure of its current status 
and of its place in the overall pattern of galaxy evolution. Kennicutt (1998) proposed
an array of star formation rate (SFR) indicators and quantitative relations for their use. 
There are two basic approaches included in his SFR metrics. One is to use optical or ultraviolet data, 
which can be traced back directly to the outputs of hot, young stars but must be corrected for 
interstellar extinction.  The second is to treat the far infrared outputs of galaxies as a 
calorimeter, so the luminosity in this spectral range is a measure of the total power being 
produced by hot, young stars. 

Recent work has combined the ultraviolet, optical, and infrared indicators 
(e.g., Iglesias-P\'aramo et al. 2006, Dale et al. 2007, Calzetti et al. 2007).
The motivation is to make the most accurate and comprehensive determination
of the SFR, assuming that a range of observations is available for a galaxy. However,
we often face a different challenge: given a very limited suite of observations, what is
the best we can do in determining SFRs? This paper addresses this problem, and
specifically the issues of: 1.) how to determine SFRs from single-band (e.g., 24$\mu$m) infrared measurements; and 
2.) under what conditions such determinations are reasonably accurate. The paper is motivated
by the success of MIPS (Rieke et al. 2004) 24$\mu$m photometry in measuring large
numbers of faint galaxies whose infrared outputs are not currently accessible at
longer wavelengths due to confusion noise and sensitivity limitations. Previous
work (e.g., Papovich \& Bell 2002; Dale et al. 2005; Smith et al. 2007) has emphasized the broad variety
of infrared spectral energy distributions (SEDs) and the resulting uncertainties
in the bolometric luminosities and thus SFRs extrapolated from 24$\mu$m measurements. 
Nonetheless, for lack of any viable alternative, such SFRs lie at the core of
most studies of distant galaxies in the thermal infrared. 

To contend with the range of behavior between 24$\mu$m and the far infrared, we will
examine two questions. Existing libraries
of SED templates (e.g., Chary \& Elbaz 2001; Dale \& Helou 2002; Siebenmorgen \& 
Kr\"ugel 2007) show a strong pattern of behavior with bolometric luminosity. The first 
question for this paper is how much the range of SED behavior can be constrained by introducing
24$\mu$m luminosity as a parameter in selecting a suitable galaxy SED. The second question
is how useful 24$\mu$m luminosity by itself is to determine SFRs. This question is
motivated by the remarkably small scatter between the extinction-corrected Pa$\alpha$
and 24$\mu$m luminosities of infrared galaxies (Alonso-Herrero et al. 2006).
We will demonstrate that SFRs can be estimated from 24$\mu$m photometry
to an accuracy of better than 0.2 dex, comparable to the accuracy obtainable with
full far infrared luminosities. 

An important part of our arguments depends on accurate SED templates for
galaxies. We include in the Appendix a description of the set of templates
used in this paper. These templates
involve both new data from {\it Spitzer} and new
methods to combine spectroscopy, photometry, and theoretical models in a
consistent way. They apply to galaxies strongly
dominated by star formation, as judged by X-ray properties, 
(e.g., Armus et al. 2007), broad SED characteristics (Farrah et al. 2003), 
and by their mid-infrared spectra
(Genzel et al. 1998). We also construct a set of average
templates for the luminosity range ($\sim 5 \times 10^9 L_\odot
< L < 1 \times 10^{13}L_\odot$). Up to $2 \times 10^{12} L_\odot$, these
templates are based directly on observations of local galaxies; above
this luminosity, they require extrapolation and have substantial
uncertainties.  Our templates have played 
important roles in a number of recent studies (Donley et al. 2007; 
Rigby et al. 2008; Seymour et al. 2008; Donley et al. 2008). 
They are provided in tables available on line.  

We introduce the templates in Section 2 (with details in the
Appendix). In Section 3, we derive SFRs in terms of 24$\mu$m photometry.
We also provide a conversion of observed flux
densities to star formation rates as a function of redshift for {\it Spitzer}
at 24$\mu$m, {\it Herschel} at 70 and 110$\mu$m, {\it WISE} at 24$\mu$m, and 
{\it JWST} in the 20$\mu$m range. We consider the radio-infrared relation
as an independent indicator of the SFR in
Section 4 and summarize our results in Section 5.

\section{SED Templates}

The Appendix describes how we assembled SED templates for eleven local LIRGs 
and ULIRGs. We also discuss how we used them to construct average templates
over the relevant luminosity range and covering the 5$\mu$m to
30 cm spectral range. In addition, we combined
the results of Dale et al. (2007) and Smith et al. (2007) to
produce a complementary set of templates at lower luminosities.
The templates are provided in online tables; we show them in Figures 1 -- 6. 

In the Appendix, we also discuss the applicability of the templates at high redshift.
Their general behavior across the observed mid-infrared (IRAC bands),
far-infrared (MIPS 24 and 70$\mu$m bands), and radio (1.4GHz) 
appears to agree with observation out at least to z $\sim$ 2. 
There is, however, a wide range of behavior of the IRAC
bands relative to the MIPS ones, presumably because at z $>$ 0.5
IRAC probes the stellar photospheric output and MIPS the
infrared excess. The former includes the contributions
of both old and young stars, while the latter is powered
primarily by the young ones. As a result, galaxies
with the same star forming rate but differing amounts
of pre-existing stellar populations will show differences
in the relative outputs in the IRAC and MIPS bands.

At z$\sim$ 2, luminous infrared galaxies tend
to have aromatic bands of strength and line profile characteristic
of less luminous galaxies locally (e.g., Sajina et al. 2007; 
Papovich et al. 2007, Pope et al. 2008, Rigby et al. 2008; Farrah 
et al. 2008). Consequently, when used with observed
24$\mu$m (rest 8$\mu$m) observations to predict intrinsic 24$\mu$m
flux densities, the templates may return values that are too 
high (see Appendix).  It is not clear at what redshift 
this shift in infrared SED behavior begins to manifest itself.
For example, at z $\sim$ 0.7, our templates span the range
of observed far infrared colors, but their behavior at high
luminosities may indicate a similar shift in infrared colors
as is seen at z $\sim$ 2.
We also show that the radio-infrared relation is preserved at z $\sim$ 2,
at least when comparing with the far infrared ($\sim$ 100$\mu$m). 
However, there are large enough uncertainties in the current
determinations of the radio-infrared relation for the MIPS 24$\mu$m
band at high redshift that the expected 
(small) shift in it cannot be verified (see Section 4). 

The indicated shifts in behavior from local to z$\sim$ 2 infrared
galaxies are modest (at most factors of two). Thus, the local
templates can give a reasonably accurate picture of the
behavior of the high redshift ones. However, since there are 
no local purely star-forming ULIRGs with luminosities above 
$\sim 2 \times 10^{12} L_\odot$, all determinations of star formation
rates at very high luminosities (including virtually all 
individually-detected galaxies at $z \sim 2$) are highly uncertain.
Further work is needed
to improve our understanding of the detailed behavior of
high redshift galaxies in the far infrared, either verifying
the use of local templates to represent them or indicating
more clearly than with present knowledge how they need to
be modified.  

\section{Infrared Determination of Star Formation Rates}

SFRs are widely estimated from far infrared observations using the
far infrared luminosities and the formulation of 
Kennicutt (1998). There are ambiguities in this approach
because the far infrared luminosity of a galaxy generally has two
components, one powered by young stars to which the Kennicutt
formulation applies, and a second, cooler component probably powered
largely by the interstellar radiation field (e.g., Devereux \& Eales 1989, Popescu et al.
2002), which should be excluded
from the Kennicutt formula. Another issue arises with {\it Spitzer}
observations because the 24$\mu$m data are much deeper
relative to a given SFR and also far less confused than the data
at 70 and 160$\mu$m. Thus, the far infrared luminosities cannot
be measured for many galaxies detected at 24$\mu$m. Other missions will face similar issues due to
confusion noise (e.g., {\it Herschel}), or will not provide 
capabilities at all relevant wavelengths (e.g., {\it WISE}, {\it JWST}). 

The usual approach to determining SFRs with single-band observations
has been to redshift a SED template to match the source, normalize it to the observed
flux density, determine L(TIR) or L$_{tot}$(IR)\footnote{Errors can be introduced into
the interpretation of far infrared data through differing definitions of the
far infrared luminosity. In this paper, in addition to L(TIR) based on 
{\it IRAS} data and as used by Sanders et al. (2003), we define
L$_{tot}$(IR) as the luminosity obtained by integrating the SED of a galaxy
from 5 to 1000$\mu$m. In addition, L(FIR) designates the portion of the far
infrared output of a galaxy that is powered by young stars. It is discussed in
Section 3.1.2.}, and iterate to match the correct
template for the estimated source luminosity. 
The accuracy of the results obviously depends
on the quality of the SED template library. 
The multiple steps in this procedure each introduce 
additional potential errors.  

To mitigate these problems, 
in this section we will first derive the relation
between 24$\mu$m flux density and the SFR. We demonstrate how the
derived relation can avoid the ambiguities in SFRs measured in the
conventional way based on total or far infrared luminosities. We 
apply our result at different redshifts, using K-corrections
from MIPS-observed to rest 24$\mu$m flux densities, based on our average templates. 
We provide the results in
the form of fits that, with interpolation, can be used for a direct
conversion of 24$\mu$m data into SFRs. The results for MIPS also
apply to {\it WISE}. We carry out
a similar calculation for the 70$\mu$m and 100$\mu$m bands of PACS on {\it Herschel} 
and for three bands near 20$\mu$m for MIRI on {\it JWST}.

\subsection{SFRs from 24$\mu$m Photometry}

\subsubsection{Relation between Pa$\alpha$ and 24$\mu$m Luminosities}

In typical star-forming regions, hydrogen recombination line strength is
a basic metric to estimate the level of massive star formation. Therefore, 
we provide an updated derivation of the
relation between L(24) and L(Pa$\alpha$) for luminous star forming regions and galaxies.
We used the data for luminous galaxies of Alonso-Herrero et al. (2006) plus 
the dataset assembled by Calzetti et al. (2007) (kindly provided
by D. Calzetti) to provide extensive coverage at lower luminosities. We applied some 
small corrections to the first data set. Where the redshift of the galaxy put the Pa$\alpha$ line far
enough from the center of the NICMOS filter to affect the transmission
by more than 5\%, we corrected the line strength to compensate. We also
determined the transformation from IRAS 25$\mu$m to MIPS 24$\mu$m flux
density by ratioing measurements of galaxies measured in common (from
the SINGS data, Dale et al. 2007 and from Engelbracht et al. 2008), 
rather than from SED models. 
We put both data sets on a common calibration basis: 1.)
we took the bandpass corrections out of the photometry reported
by Engelbracht et al. (2008); and 2.) we corrected all pre-2007
photometry to the current MIPS calibration (Engelbracht et al. 2007). 

The resulting relationship between extinction-corrected L(Pa$\alpha$) and
L(24) for the galaxies and HII regions 
is shown in Figure 7. We have fitted the data for the $\sim$ solar ("high":
Calzetti et al. 2007)
metallicity cases with a linear relation and find a rms scatter of 0.27 dex. 
%The scatter is dominated by a small number of
%objects; if we reject the two high and two low outliers and only fit for
%$L(24) > 10^8 L_\odot$, we obtain
The scatter is dominated by two low-lying galaxies.  One of them, NGC 4736,
is in a post-starburst phase (Walker et al. 1988; Taniguchi et al. 1996),
which explains its low level of Pa$\alpha$ emission.  The other, IC 860,
has very strong H$\beta$ absorption (Kim et al. 1995), suggesting a similar
explanation.  We reject these two low outliers and the two highest outliers
and fit only for $L(24) > 10^8 L_\odot$, to obtain:

\begin{equation}
\label{eq1}
\log (L(Pa\;\alpha ))=(-1.081\pm 0.197)+(0.849\pm 0.021)\log (L(24))
\end{equation}

\noindent
with a scatter of 0.21 dex. We use the trimmed result in the following. 
The way it has been determined, it should be used with the direct
pipeline output for the MIPS 24$\mu$m photometry (no photometric bandpass
corrections). 

Our derived relation agrees with those of both Alonso-Herrero et al. (2006) and Calzetti
et al. (2007) within the errors but has smaller errors because it combines both 
of their samples into a single relation. It also has a somewhat different slope
that results largely from reconciling all the 24 and 25$\mu$m photometry to
a common basis. The scatter in this relation is similar 
to the scatter just in computing L(TIR) from different data sets. Therefore, 
using the conventional approach of estimating L(TIR) from L(24) and then
the SFR from L(TIR) interposes a step that adds substantial uncertainty
to the final result without improving its accuracy.
   
The relationship between L(Pa$\alpha$) and L(24) is not 
proportional, but shows a progressively greater output
at 24$\mu$m relative to Pa$\alpha$ with increasing luminosity. 
It is well known that the extinction in star-forming galaxies is
greater at higher SFR (e.g. Wang \& Heckman 1996; Buat et al. 2007) and that it may
require empirical correction to reconcile Balmer-$\alpha$ derived
SFRs to radio or infrared measurements (e.g. Sullivan et al. 2003).  The
present work extends this problem to Pa$\alpha$ and to higher infrared luminosities.
There are two
possibilities. First, the HII regions in more luminous galaxies may raise the
dust to a higher temperature and cause additional emission at 24$\mu$m relative
to the Pa$\alpha$ luminosity (Calzetti et al. 2007). Alternatively, at high luminosities
there may be a larger proportion of young massive stars embedded in 
ultra-compact HII regions, from which Pa$\alpha$ cannot escape 
and within which the dust may absorb a significant fraction of the ionizing photons
(Rigby and Rieke 2004, Dopita et al. 2006); in this 
case, extinction-corrected Pa$\alpha$ may underestimate the SFR and the
24$\mu$m luminosity may be a better indicator.  
Consistent with this possibility, in the young star forming system C of Arp 299, 
Alonso-Herrero et al. (2008) find a deficiency of flux
in the mid-infrared fine structure neon lines, consistent with a
substantial portion of very heavily embedded and dense star forming regions. 
 
As a test of these possibilities, we look at the implications of setting the SFR proportional
to the extinction-corrected Pa$\alpha$. We use the relation
between L(24) and L(TIR) from the Appendix and anticipate the results of the
following sections to derive under these assumptions:

\begin{equation}
\label{eq2}
sfr=6.57\times 10^{-10}\;L(TIR)^{0.898\pm 0.022}
\end{equation}

\noindent
Here and in the following, we indicate intermediate estimates of
the star formation rate in lower case to distinguish them from
final formulations.  The deviation from linearity in Equation 2 
(slope $\neq 1$) would imply that over the range of about three orders
of magnitude in the luminosities of starbursts, LIRGs, and ULIRGs,
the generation of bolometric infrared luminosity from young stars
varies in efficiency by a factor of two. Since there is no
easy physical explanation, we instead adopt the point of
view that the SFR is proportional to the appropriately calculated
infrared luminosity. This assumption is identical to that
made by Kennicutt (1998) in formulating his widely-applied
relationship between SFR and L(FIR).  Simply stated, we (and
Kennicutt) assume that the FIR acts as a calorimeter, that
virtually all of the luminosity of the young stars is captured
and re-emitted in the FIR.

There is an important additional implication of the non-proportionality between Pa$\alpha$ and
L(TIR). It suggests that it may be
impossible to capture the full hot stellar output in very luminous galaxies using 
optical and near infrared hydrogen recombination lines. That is, SFRs estimated on
the basis of extinction-corrected Pa$\alpha$
should be taken as lower limits, and for very luminous galaxies,
lower limits by significant factors. This issue will grow in
importance with the use of recombination lines (such as H$\alpha$) at shorter
wavelengths than Pa$\alpha$.

\subsubsection{Calibration of the SFR via Far Infrared Luminosity}

Kennicutt (1998) has calibrated the SFR in terms of bolometric
luminosity, which he assumed emerges in the FIR:

\begin{equation}
\label{eq3}
SFR_{FIR} (M_\odot \;yr^{-1})=4.5\times 10^{-44}\;L_{FIR} (ergs\;s^{-1}),
\end{equation}

\noindent
This formula was derived from theoretical starburst models and the 
calorimetric assumption that all their luminosity would emerge in the FIR
(R. Kennicutt, private communication, 2008).
Applying the formula is not as straightforward as it appears
because it is difficult to define observationally the correct form
of L$_{FIR}$. The total
infrared luminosity, L$_{tot}$(IR), includes
the very far infrared and submm, spectral regions that in starburst-luminosity
galaxies tend to be dominated by the emission of cold dust\footnote{Our notation
of L$_{tot}$(IR) distinguishes this quantity, obtained by integrating
the galaxy output from 5 to 1000$\mu$m, from L(TIR), which is
calculated according to the procedure recommended by Sanders et al. (2003)
on the basis of the {\it IRAS} measurements alone.}. This dust
is probably heated by the interstellar radiation
field dominated by an older population of stars, not by the hot, newly formed stars (e.g., 
Devereux \& Eales 1989; Popescu et al. 2002).
In addition, there are different algorithms for computing
L$_{tot}$(IR), increasingly used in place of L(FIR) as specified by Kennicutt. 
They produce somewhat different answers from similar data.
Finally, the far infrared SEDs of galaxies are not
usually well sampled and in many situations (e.g., deep cosmological
surveys) are technically very difficult to sample to great depth because
of confusion noise. 

To apply Kennicutt's relation between SFR and L(FIR) accurately 
requires that we isolate the portion of L$_{tot}$(IR) that is powered
by the young stars. 
Figure 6 suggests a simple way to do so. 
Our lowest-luminosity templates have a substantial contribution
from cold dust, but as the luminosity is increased the far infrared
regions converge to a single form. The templates show that
the more vigorously star forming galaxies are producing a single
unique far infrared SED that eventually overwhelms
the cold dust component (see Devereux \& Eales (1989) for a discussion
of this decomposition of far infrared SEDs). 

To make use of this behavior, we quantify it in terms of L(24)/
L$_{tot}$(IR) computed from the templates and shown in Figure 8. 
The initial increase with luminosity in the proportion of the luminosity emerging at 
24$\mu$m results from the increasing
prominence of the emission by warm dust heated by young stars,
over the cold dust heated by the interstellar radiation field. 
At very high luminosity, there is a modest decrease in the
ratio probably due to increased optical depth in the star forming
regions. We want to avoid these effects in our calibration of the
SFR, so we will use the peak ratio at log(L(TIR))=11, 
L(24)/L$_{tot}$(IR) = 0.158. From Figure 8, the
smallest plausible value is 10\% lower. This result allows us to modify Equation 3 
so it applies to measurements at 24$\mu$m. If we also put the luminosities 
into solar units, we have

\begin{equation}
\label{eq4}
sfr_{FIR} (M_\odot \;yr^{-1})=1.09 ~\rm{to}~ 1.20 \times 10^{-9}\;\rm{L(24\mu m,\;L_\odot)} ,
\end{equation}

\noindent
where the range corresponds to the 10\% range of L(24)/L$_{tot}$(IR)
discussed above. Because 24$\mu$m is in the heart of the spectral range dominated by the
warm dust associated with recent star formation, the new formulation
circumvents the issues in the original formula associated with the
definition of L(FIR). 

The infrared metrics for SFR are based on a calorimetric argument, and
they are subject to systematic errors if the calorimeter is leaky, i.e. if
significant amounts of the stellar luminosity escape directly in the
ultraviolet rather than being absorbed and reradiated in the infrared. 
This issue can be quantified by comparing L(UV) with L(TIR). Various
studies (e.g., Bell 2003; Schmitt et al. 2006; Buat et al. 2007)
have evaluated this behavior. For infrared-selected galaxies, they 
agree that the UV contribution to the total young-star-powered
luminosity is only about 20\% at log(L(TIR)) $\sim$ 9.75 and decreases
rapidly with increasing infrared luminosity, e.g. to $\sim$ 8\% at
log(L(TIR)) $\sim$ 10.5. Therefore, the calorimeter 
assumption is questionable for log(L(TIR)) $<$ 9.5, but good for 
log(L(TIR)) $\ge$ 9.75. At 10$^{11}$ L$_\odot$, the average level of
leakage is only $\sim$ 2.5\% (Buat et al. 2007). If we correct
Equation 4 for this effect, we get

\begin{equation}
\label{eq5}
sfr_{FIR} (M_\odot \;yr^{-1})=1.12 ~\rm{to}~ 1.23 \times 10^{-9}\;L(24\mu m,\;L_\odot 
).
\end{equation}

\subsubsection{Calibration of the SFR via Hydrogen Recombination Lines}

Although the above derivation is simple in concept, it relies on a secondary
indicator for the SFR, namely the far infrared output associated with the
absorption of the young stellar luminosity by interstellar dust. A
more direct metric for the SFR can be based on hydrogen recombination
lines, excited directly by the young, hot stars. 
A strong connection between the hydrogen recombination lines and L(24) is indicated both 
by the small scatter in the fit of L(Pa$\alpha$) vs. L(24),
and by the spatial correlation between the 24$\mu$m output of galaxies 
with both the HII regions and with the diffuse H$\alpha$ (e.g., Hinz et al. 2004;
Tabatabaei et al. 2007). We therefore explore a calibration of the 
SFR in terms of these lines.

We map L(Pa$\alpha$) to L(24) by constraining the
fit in Figure 7 to $7 \le$ log(L(24)) $\le 10$, corresponding
roughly to $8 \le$ log(L(TIR)) $\le 11$. We assume over this
range that the optical depth effects that arise at very high
luminosity are not strong. Also, although we fit down to the
luminosity range where the luminosity escaping in the UV is large for
whole galaxies, much of the data in this range is for individual HII
regions within large, metal-rich galaxies so these effects are
greatly reduced. The best fit to the data reproduces
a slope similar to that for the whole range of luminosities, but
if we constrain the slope to be unity $\chi^2$ grows by only
14\%. The result of this latter case is that

\begin{equation}
\label{eq6}
L(Pa\;\alpha )=0.038\;L(24).
\end{equation} 

\noindent
This calibration is identical to that derived by Kennicutt et al. (2007)
for the HII regions in M51. Although these data are included in our fit,
we have also included full galaxies as well as additional HII regions.  

The nebular conditions in starbursting galaxies are well constrained 
(e.g., Roy et al. 2008 and references therein), with low electron temperatures
($\sim$ 5000K) and densities of 500 - 50000 cm$^{-3}$. The range of case B
recombination line intensities is small under these conditions. 
The relation proposed by Kennicutt (1998) 
can be converted to use with Pa$\alpha$ assuming a ratio of 
P$\alpha$/H$\alpha$ = 0.128 (Hummer \& Storey 1987): 

\begin{equation}
\label{eq7}
SFR_ (M_\odot \;yr^{-1})=6.2\times 10^{-41}\;L(Pa\alpha ,\;ergs\;s^{-1}).
\end{equation}

As indicated by Alonso-Herrero et al. (2006), Kennicutt et al. (2007), and
Calzetti et al. (2007), it is possible that the calibration
applies poorly to whole galaxies because of diffuse extended
Pa$\alpha$ emission that is not captured by the NICMOS imaging
used to measure the Pa$\alpha$ line strength. We can set an upper
limit on such diffuse emission using SFR indicators that are
sensitive to it, such as H$\alpha$ imaging, deep 24$\mu$m imaging, 
or filled-aperture high frequency radio photometry. M33 has
been thoroughly studied in all three of these indicators
(Devereux et al. 1997; Hinz et al. 2004; Tabatabaei et al. 2007). 
Of order 30\% of the free-free, H$\alpha$, and 24$\mu$m signals
are associated with a diffuse component. The extinction
to this component is very small, whereas that for the compact
sources is A$_V$ $\sim$ 1, so H$\alpha$ images 
overemphasize the relative intrinsic strength of the diffuse emission.
The largest body of applicable data for other galaxies is
H$\alpha$ imaging (e.g., Devereux et al. 1994; Devereux et al.
1996; Hameed \& Devereux 2005). These studies show that
30\% is an approximate upper limit for the fraction of
diffuse H$\alpha$ and indicate that it should
be much less obscured than the emission associated with
discrete sources. We conclude that the relative intrinsic level of
diffuse emission associated with recent SFR is on average
no more than $\sim$ 15\%.

Making use of Equation 7, we find

\begin{equation}
\label{eq8}
sfr_{Pa\,\alpha } (M_\odot \;yr^{-1})=1.15 ~\rm{to}~ 1.30\times 10^{-9}\;L(24\mu m,L_\odot 
)
\end{equation}

\noindent
where the range is without (1.15) and with (1.30) an allowance for
diffuse emission. This expression is virtually identical to sfr$_{FIR}$ (Equation 5). 
The agreement of the two independent estimators implies that the systematic 
errors are well controlled. The combined result from Equations 5 and 8
under the Kennicutt guidelines is then

\begin{equation}
\label{eq9}
sfr_{Kennicutt} (M_\odot \;yr^{-1})=1.18\times 10^{-9}\;L(24\mu m,L_\odot ).
\end{equation}
\subsubsection{Initial Mass Function}

The derived SFR is based only on the outputs of very massive stars
and hence provides virtually no constraint on the rate of formation
of low mass stars. However, the quoted rate for the total formation of stars
conventionally integrates down to the minimum stellar mass, $\sim$ 0.1 M$_\odot$, 
and therefore depends strongly on the assumed initial mass function
(IMF). Kennicutt (1998) assumed a Salpeter IMF with a 
single power law slope of -1.35 from 0.1
to 100M$_\odot$. However, Rieke et al. (1993) and Alonso-Herrero et al. (2001) show
that such an IMF violates plausible constraints on the dynamical mass
for the starbursts in M82 and NGC 1614. 

Rieke et al. (1993) derived forms of the IMF {\it ab initio} for M82 that illustrate
some general requirements.  They argued
that the IMF in M82 needed to have relatively more massive stars
{\it than the IMFs then proposed for the local field} (e.g., Miller
\& Scalo 1979; Basu \& Rana 1992). All of these
local IMFs fell toward high masses much faster than the Salpeter
slope. We can compare various forms of IMF with roughly similar slopes by comparing
the mass in stars above 10 M$_\odot$. IMF8, 
the formulation favored in the models
of Rieke et al. (1993), has a net slope at high mass similar to the Salpeter value
and produces the identical proportion
of such stars as the IMF proposed by Kroupa (2002), which has a slope
of -0.3 from 0.08 to 0.5 M$_\odot$ and of -1.3 from 0.5 to 100 M$_\odot$. 
A similar result applies to the Chabrier (2003) IMF. 
The widespread adoption of the Salpeter-like slope 
with a more shallow slope at low masses to fit extragalactic
star forming regions is therefore a confirmation 
of the results of Rieke et al. (1993).
The total mass for any of these IMFs is $\sim$ 0.66 times that of
the unbroken Salpeter form that yields the same mass in stars
$>$ 10M$_\odot$ (as was adopted by Kennicutt (1998)). 

Rieke et al. (1993) also showed that the formation
of extremely massive stars should not be too strongly favored or an embarrassingly
large amount of oxygen will be produced (see also Wang \& Silk 1993).
This constraint probably eliminates IMFs with slopes flatter than the
Salpeter value (Gibson 1998).

Therefore, we have a final form for the SFR:

\begin{eqnarray}
\label{eq10}
SFR(M_\odot \;yr^{-1}) =7.8\times 10^{-10}\;L(24\mu m,L_\odot ) \nonumber \\ 
=0.66\;SFR_{Kennicutt},
\end{eqnarray}

\noindent
for $5 \times$ 10$^9$ L$_\odot$ $\le$ L(TIR) $\le$ $1 \times$ 10$^{11}$ L$_\odot$ 
or $6 \times$ 10$^8$ L$_\odot$ $\le$ L(24) $\le$ $1.3 \times$ 10$^{10}$ L$_\odot$,
where L(24) is in the rest frame and is as measured with MIPS with no bandpass corrections. 
For L(24) $>$ $1.3 \times 10^{10}$ L$_\odot$, 

\begin{eqnarray}
\label{eq11}
SFR(M_\odot \;yr^{-1})=7.8\times 10^{-10}\;L(24\mu m,L_\odot ) \nonumber \\
\times
(7.76\times 10^{-11}L(24\mu m,L_\odot ))^{0.048}.
\end{eqnarray}.

\noindent
The final term accounts for the slight decrease in L(24)/L(FIR) with increasing
luminosity above L(TIR) = 10$^{11}$ L$_\odot$ (see Figure 8). 
This equation only holds up to $2 \times 10^{12} L_\odot$, since there are 
no local star forming ULIRGs to constrain the templates above this luminosity.

Determining L(24) also allows selection of an appropriate SED template. 
The K-correction to the observed 24$\mu$m flux density is computed from this 
template and the redshift. 
In this manner, SFRs can be estimated solely from 24$\mu$m observations for
dusty, luminous star-forming galaxies at any redshift. 

\subsubsection{Uncertainties}

The lack of knowledge of the low mass IMF in luminous star forming galaxies is
probably the dominant uncertainty in calculating the SFR. We have brought
the formulation into agreement with estimates of the local IMF (Kroupa 2002, Chabrier 2003)
and other constraints, but it remains possible that low mass stars
are significantly less common in vigorously star forming regions than
implied by these IMFs. 

Because the original formulation by Kennicutt (1998) has been so widely used,we have made no other changes to its input parameters. That is, we have
adopted his modeling to convert SFRs to luminosities in the hydrogen
recombination lines and in the far infrared.  This approach allows updates in the theoretical
modeling to be applied in a straightforward way and also allows work
using the original Kennicutt (1998) formula to be adjusted unambiguously.
We therefore consider only the uncertainties in mapping the predicted
SFR metrics into L(24). 

The calibration of the SFR in terms of L(FIR) is subject to
two types of error. The first is in the conversion of
L(FIR) to L(24). An approximate measure of this error is the
scatter in the relation between L(24) and L(TIR) (see Appendix), which
is 0.13 dex. The second source of error is due to the simple calorimeter
assumption that underlies the formula. 
Some of the luminosity of the young stars escapes in the UV.
Although we have accounted for this effect on average, there is 
a substantial variation from one galaxy to another.
Figure 7 in Buat et al. (2007) indicates a 
1-$\sigma$ scatter in L$_{IR}$/L$_{UV}$ of $\sim$ 0.4 dex.
There are two contributors: 1.) variations from galaxy to galaxy in the amount of
UV escaping; and 2.) variations due to the anisotropy of the
escaping UV radiation (e.g., less will escape along the plane than
perpendicular to it for a spiral galaxy). Because the second
contribution is not relevant for the calorimeter argument, the
observed scatter provides an upper limit to the resulting uncertainty
in the SFR. This upper limit approaches 0.15
dex at log(L(TIR) = 9.75 but falls below 0.1 dex for
log(L(TIR) $\ge$ 10.  

For the calibration of SFR based on L(Pa$\alpha$), there is
a scatter of 0.27 dex relative to the fit where we have
constrained the slope between L(Pa$\alpha$) and L(24) to
be unity. This scatter becomes smaller toward
higher infrared luminosities (see Figure 7), so in many
applications 0.27 dex can be taken as an upper limit.

Given the agreement of our two independent estimates of the SFR
along with the individual uncertainties, we conclude that 
Equations 10 and 11 should be accurate to within 0.2 dex. This
estimate omits uncertainties due to the IMF and those
within Kennicutt's (1998) original derivation of the theoretical
relation between the SFR and L(FIR) or L(H$\alpha$).

\subsection{Practical Applications: Estimating Star Formation Rate and Infrared Luminosity}

\subsubsection{{\it Spitzer} at 24$\mu$m}

To illustrate our approach to determining SFRs, 
we use the template SEDs to determine the relation between
the observed and rest 24$\mu$m flux densities for star forming
galaxies over a range of redshifts. The rest flux densities can be used
with the relations above to determine SFRs from the same observations.
We provide calibrations for SFR as a function
of redshift and observed flux with Spitzer 24 $\mu$m measurements, 
and for future infrared observatories.

Each of the average template SEDs constructed in Section 6.2 has a value 
of $L(TIR)$, from which we computed the corresponding
values of $L_{24}$ restframe and SFR using Equations 25 and 11.  We then 
convolve the SED with instrumental response curves to compute 
K-corrections, $K_{corr}$ in dex, from the observed IR band $\nu_{obs}$
to the restframe 24 $\mu$m, to predict the luminosity in the observed band.
Assuming a cosmology yields the luminosity distance $D_L(z)$ and the observed flux.
For $L_\nu \propto \nu f_\nu$,  the K-correction to rest 24 $\mu$m is:

\begin{equation}
\label{eq12}
K_{corr} = \rm{log}[ (1+z) \frac{f_{\nu}(\nu=(1+z)\nu_{obs})}{f_\nu(24 \mu m)}]
\end{equation}

%\begin{equation}
%\label{eq22b}
%\frac{L_{\nu,obs}}{\nu_{obs}} = \frac{L_{24,rest}}{\nu_{24}} 10^{-K_{corr}},
%\end{equation}

\begin{equation}
\label{eq13}
4 \pi {D_L}^2 f_{\nu,obs} = \frac{L_{\nu,rest}(24 \mu m)}{\nu_{24}} 10^{K_{corr}}.
\end{equation}

\noindent
This procedure yields tracks of the template
SEDs in $4 \pi {D_L}^2 f_{\nu,obs}(z)$,
shown in Figure 9 for observations at 24 $\mu$m.  
We find that at a given redshift, the dependence of SFR on $f_{\nu,obs}$
is closely approximated by a power law.  There is a small kink
in the relation at $L(TIR) \sim 10^{11}$ L$_\odot$ that deviates from the
power law by at most 0.1 dex.  The slope and intercept of the
power law vary with redshift as the 24 $\mu$m band probes
different regions of the infrared SED.  At higher redshifts,
the slope of SFR on observed 24 $\mu$m flux is steeper.

To derive a simple recipe for
estimating SFR from $f_{24,obs}$, we fit power laws to the SFR--$f_{24,obs}$
relation and tabulate the results below.
The power laws are parametrized by intercept $A$ and slope $B$,
and zeropointed at ${\rm log} (4 \pi {D_L}^2 f_{24,obs}) =53$
to reduce covariance in the fit parameters:

\begin{equation}
\label{eq14}
{\rm log (SFR)} = A(z) + B(z) \times ({\rm log} (4 \pi {D_L}^2 f_{24,obs}) - 53),
\end{equation}

\noindent
where SFR is in $M_\odot \rm{yr}^{-1}$ and $4 \pi {D_L}^2 f_{24,obs}$ is in Jy~cm$^2$.  
Figure 10 shows the trends of $A$ and $B$ with redshift for observations at 
24 $\mu$m (heavy black line) and for other infrared filters, and the values
of $A(z)$ and $B(z)$ are given in Tables 1 and 2.  Silicate absorption
and its dependence on luminosity introduce a feature in $A(z)$ and $B(z)$,
at $z=1.4$ for 24 $\mu$m.  

To apply this relation given a redshift and flux $f_{\nu,obs}$,
the reader should determine the values of $A$ and $B$
by interpolation in redshift in Table 1 or 2, and multiply $f_{\nu,obs}$ by
$4 \pi {D_L}^2$ in cm$^2$ for the chosen cosmology.
Tables 1 and 2 provide the coefficients $A(z)$ and $B(z)$ for
several infrared filters, including {\it Spitzer}/MIPS 24 $\mu$m, {\it Herschel}/PACS 70 $\mu$m
and 100 $\mu$m, and {\it JWST}/MIRI filters at 18, 21, and 25 $\mu$m.
The Spitzer/MIPS 70 $\mu$m and the {\it WISE} 23 $\mu$m filters are
very close to PACS 70 $\mu$m and Spitzer/MIPS 24 $\mu$m respectively,
and the corresponding values in Table 1 can be used.
Total infrared luminosity can be estimated from 
a flux measurement in one of these filters by using $A(z)$ and $B(z)$ to
derive SFR via Equation 14, and then applying Equations 11 and 25 to
yield $L(TIR)$.

\section{Infrared/Radio Relation}

An alternative extinction-free approach to estimating SFRs for
infrared galaxies is to utilize the proportionality 
between the radio and infrared emission of galaxies originally
found by van der Kruit (1971) and Rieke \& Low (1972) and shown to be universal
with {\it IRAS} (Helou et al. (1985)). In this section, we calibrate
this relation consistently with the preceding work on L(24). 

\subsection{Infrared/Radio Relation for Local Galaxies}

The {\it IRAS} data were primarily used to study the infrared-radio relation in the far infrared,
using the 60 and 100$\mu$m bands. With the high sensitivity of {\it Spitzer}, interest
has grown in determining and testing the relation at 24$\mu$m. 
The first such study, by Appleton et al. (2004), found log(f$_\nu$(24$\mu$m)/f$_\nu$(1.4GHz))= q$_{24}$ = 0.94 to 1.00 
(depending on the subsample and correction method for SED behavior) with a 
dispersion of about 0.25 dex and q$_{70}$ = 2.15 with a dispersion of 0.16 dex. 
Other studies have also determined q$_{24}$ from deep Spitzer observations. 
For example, Boyle et al. (2007) found q$_{24}$ = 1.39 $\pm$ 0.02 from stacking deep radio data 
at the positions of SWIRE sources. In comparison, Beswick et al. (2008) used the deep radio and infrared
data in the HDF to derive q$_{24}$ = 0.52. Ibar et al. (2008) find a K-corrected value out 
to z $\sim$ 3 of q$_{24}$ = 0.71 with a dispersion of 0.47. 
Marleaux et al. (2007) obtained results consistent with those of Appleton et al. (2004). 
Gruppioni et al. (2003) measured the analogous q$_{15}$ with ISO data, obtaining a value 
of $\sim$ 0.82 but estimating that the average ratio must be 
adjusted upward by a factor of two, i.e., q$_{15}$ $\sim$ 1.12 with a correction for incompleteness 
in the radio. The equivalent q$_{24}$ would be $\sim$ 1.2 not corrected for incompleteness and 
$\sim$ 1.5 when corrected (based on the 15 to 24$\mu$m colors of our templates).

Clearly, these estimates are not in good agreement. Evidently the selection effects,
K corrections, and other issues in determining q$_{24}$ using faint survey data have undermined
some of the estimates and caused the set to diverge. Therefore, we will determine q$_{24}$ for
nearby galaxies using high signal to noise measurements from the {\it IRAS} bright galaxy 
sample (BGS; Sanders et al. 2003) and the VLA. This sample is selected to have 60$\mu$m 
flux density greater than 2Jy and virtually all members have reliable measurements at both
25 and 60$\mu$m (at least). Given the infrared-radio relation, the 60$\mu$m threshold corresponds on 
average to a flux density at 1.4GHz of about 12 mJy, which is well within the detection range of the VLA 
even for short exposure surveys. We used primarily radio data from Condon et al. (1996) and Condon et al. 
(1998), augmented in a few cases by Condon et al. (1983), Condon et al. (1991), Condon et al. (2002), Iono et al. 
(2005), and Baan \& Kl\"ockner (2006).  We eliminated galaxies with log(L(TIR)) $<$ 9.5 because the 
infrared outputs may not be a reliable measure of the star formation rates (SFRs) at such low luminosities
We also eliminated galaxies with AGN, based on the studies of Maiolino \& Rieke (1995), Ho et al. (1997), 
and Veilleux et al. (1995), plus a few objects we identified as outliers in q$_{24}$
and that we found were known AGN. 
Our final sample consists of 373 galaxies, well over half of the original BGS. The largest number of 
rejections was from lack of radio data, followed by presence of an AGN, followed by IRAS data indicated 
to be of marginal quality in the BGS, followed by too low a luminosity. 

To compute q$_{24}$, we converted {\it IRAS} 25$\mu$m data to be equivalent to MIPS photometry at 24$\mu$m in two steps:
1.) we use the measurements of Dale et al. (2007) and Engelbracht et al. (2008) to determine an 
average ratio of IRAS 25$\mu$m to MIPS 24$\mu$m flux densities of 1.16 $\pm$ 0.02, for 
galaxies of average log(L(TIR)) between 10.5 and 11; and 2.) we use our templates to determine the luminosity
dependence of this correction, which ranges from 1.10 at log(L(TIR))=10 to 1.22 at log(L(TIR))=12. We 
also used the slope from the templates and an assumed slope of -0.7 for the radio spectra to compute a 
K correction for each galaxy (this correction was never more than 0.06 dex). Figure 11 shows the values 
of q$_{24}$ plotted as a function of log(L(TIR)). The average slope is -0.051 $\pm$ 0.034 for log(L(TIR))  $<$ 11, 
not significantly different from zero. We therefore use the average value for log(L(TIR))  $<$ 11, 

\begin{equation}
\label{eq15}
q_{24}=1.22 \pm 0.02.
\end{equation}

\noindent
For log(L(TIR)) $>$ 11, we obtain

\begin{equation}
\label{eq16}
q_{24}=(-1.275 \pm 0.756)+(0.224 \pm 0.066)log(L(TIR)).
\end{equation}

\noindent
The slope is significant at the 3-$\sigma$ level. The rms dispersion around this broken straight line fit 
is 0.24 dex. Taking a simple average of all the values at 60$\mu$m, we find q$_{60}$ = 2.13, with a dispersion 
of 0.21 dex. Using the Yun et al. (2001) definition of FIR, we obtain q$_{FIR}$ = 2.42 with a dispersion of 0.23, 
compared with their value (for a different sample that includes ours) of 2.34.

Our value at 24$\mu$m is substantially larger than many of those based on deep {\it Spitzer} surveys. 
One possibility for these discrepancies would be if the radio flux densities in our sample are 
systematically underestimated. The most plausible cause would be missing baselines in the VLA images. 
However, the great majority of the radio flux densities were obtained from Condon et al. (1996) and 
Condon et al. (1998), where care was taken to provide data at small baselines (and with large 
synthesized beams, $\sim$15" and 45", respectively). We nonetheless tested for underestimation by
comparing the flux densities we used with those from the Green Bank survey (Becker \& White 1992),
based on filled aperture observations with a 700" diameter beam. After eliminating sources with
cataloged bright confusing radio sources in the beam and three more where the large flux
densities in the GB data implied uncataloged confusing sources, we were able to include
61 galaxies from the BGS in this comparison. We found that on average they were 7 $\pm$ 3\% brighter
in the GB data than for the values we adopted, a difference of only 0.03 dex. 
An over-estimate of the IRAS fluxes does not seem plausible. Nonetheless, to test for one,
we repeated our calculations using the MIPS data on the SINGS sample tabulated by Dale et al. (2007),
finding q$_{24}$ = 1.30 and (for only 14 galaxies) a ratio of 1.086 for the fluxes measured in
the GB survey divided by the tabulated ones. That is, the agreement with the results
from the BGS is very good.

However, a number of the galaxies
in Dale et al. (2007) have recently been measured in the radio 
at Westerbork (Braun et al. 2007) and for these measurements the discrepancy 
with our adopted values is larger, 23 $\pm$ 6\%. Galaxies with low surface brightness in the radio appear 
to dominate this discrepancy. Given that the large beam GB survey should capture this flux, 
we place more reliance on that comparison and conclude that
the radio could be only slightly ($\sim$ 7\%) underestimated in our study.

Although the {\it IRAS} data were seldom analyzed for the radio-infrared 
ratio at 25$\mu$m, the ratio of 60$\mu$m to 25$\mu$m flux densities for star forming galaxies is 
typically slightly less than ten (e.g., Rieke \& Lebofsky 1986). 
Our derived value for q$_{24}$ is consistent 
with this constraint, but the substantially smaller values of q$_{24}$ 
(e.g., Beswick et al. 2008) are not. In addition, 
any direct conflict with the deep survey results is not well established. Extrapolation 
back to z $\sim$ 0 by means of Figure 8 in Beswick et al. (2008) or Figure 14
in Marleau et al. (2007) indicates that their works may actually imply a similar value to
ours at z = 0, but with large internal errors. 

Given the indication that we slightly overestimated q$_{24}$ from the comparison
with the filled aperture radio fluxes, but that we might have slightly
underestimated it relative to the results from Dale et al. (2007), we elect
to accept the BGS result without adjustment. The resulting estimator of the
SFR 

\begin{equation}
\label{eq17}
SFR(M_\odot \;yr^{-1})=1.13\times 10^{-4}\;\rm{L(1.4\; GHz,L_\odot)}
\end{equation}

\noindent
for $5 \times$ 10$^9$ L$_\odot$ $\le$ L(TIR) $\le$ $1 \times$ 10$^{11}$ L$_\odot$ 
or $4 \times$ 10$^3$ L$_\odot$ $\le$ L(1.4 GHz) $\le$ $8.7 \times$ 10$^{4}$ L$_\odot$,
where L(1.4 GHz) is in the rest frame. 
For L(1.4 GHz) $>$ $8.7 \times 10^{4}$ L$_\odot$, 

\begin{eqnarray}
\label{eq18}
SFR(M_\odot \;yr^{-1})=1.13\times 10^{-4}\;\rm{L(1.4\; GHz,L_\odot )}\quad \nonumber \\
\times (1.15\times 10^{-5}\;\rm{L(1.4\; GHz,L_\odot )})^{0.27}.
\end{eqnarray}

\noindent
Given the scatter in q$_{24}$, we estimate that these relations may be
accurate to about 0.35 dex.

\subsection{Redshift Dependence of the Infrared-Radio Relation}

We now compare the radio-infrared relation 
with observations of galaxies
at high redshift to test whether there is evolution. 
Figure 12 (reproduced from Seymour et al. 2008) 
explores the relation at z $\sim$ 2. It shows the templates of
local ULIRGs dominated by star formation and with measurements
at 1.4 and 8GHz. At ULIRG luminosity (and for the even more luminous
z $\sim$ 2 galaxies), the entire far infrared SED should be dominated
by the power from young, hot stars (see Figure 8). 
The SEDs are normalized at 260$\mu$m. In addition,
the figure shows the 1.4GHz and submm measurements
of luminous infrared galaxies at z $\sim$ 2 by Kov\'acs et al. (2006)
(including only cases with z $\ge$ 1.4 and excluding numbers 2, 3, 9,
and 10 in their Table 1, since three of them appear to be significantly
contaminated by AGN and one was not detected at 350$\mu$m).
When possible, we used the radio reductions of Biggs \& Ivison (2006) 
in preference to those of Kov\'acs et al. (2006). 
For each high-redshift galaxy, the fluxes were scaled to provide
a good fit to the templates at 
%The high-redshift
%measurements of each galaxy were normalized to provide a good fit at
both 350 and 850$\mu$m in the observed frame. The results imply
that the radio and infrared flux densities of the high redshift galaxies
are in good agreement with the local templates, although with large
scatter. It is possible that the scatter is increased by
measurement errors. In any case, there is no apparent offset.

We used the measurements in Figure 11 to determine a relation between 
redshift and the ratio of flux densities at 850$\mu$m and 1.4 GHz for 
galaxies behaving like the local ULIRGs. We did so by first averaging 
(in the logarithm) the spectral templates in the far infrared and 
submm, and also the radio observations at 1.4 and 8.4 GHz (Condon et 
al. 1991). We fitted the two radio points with a power law, and also 
fitted a power law between 180 and 380$\mu$m (approximately the rest 
spectral range of the 850$\mu$m data at the redshifts of interest). 
These fits allows us to derive

\begin{equation}
\label{eq19}
\frac{{S\left( {850\mu m} \right)}}{{S\left( {1.4GHz} \right)}} = 
82.4\left( 
{\frac{{z + 1}}{{3}}} \right)^{3.15}.
\end{equation}

We applied this fit to all of the sources with z $>$ 1.4 in Chapman et 
al. (2005) measured at least to 4-$\sigma$ at 850$\mu$m; to the sources  
Lock 850.01, 850.03, 850.12, 850.17, 850.18, and 850.33 from Ivison et 
al. (2007), that is those with redshifts $>$ 1.4 and without AGN 
components in the radio; and to the sources at z $>$ 1.4 in Kov\'acs et 
al. (2006). The total is 39 sources. We found that the average 
deviation from Equation 19 was only 13\% $\pm$ 10\%, in the sense that the 
local ULIRGs have slightly brighter relative radio outputs than the 
high-z ones. The uncertainties are likely to be understated by the 
nominal error, in part because of the limited number of local ULIRGs we 
have used and in part because the measurements at high redshift have 
substantial, perhaps understated, uncertainties. However, the 
conclusion that the radio-IR relationship does not change over the 
redshift range of z = 0 to z $\sim$ 3 is strongly indicated. 

The lack of evolution of galaxy SEDs as indicated in Figure 11
validates using submm and radio data as a photometric redshift
indicator (e.g., Hughes et al. 2002, Aretxaga et al. 2005).

The lack of evolution of the radio-infrared relation has also been
demonstrated by, e.g., Ibar et al. (2008), at observed 24$\mu$m, 
although some other investigators claim to see low levels of evolution
(e.g., Kov\'acs et al. 2006; Vlahakis et al. 2007). 

\section{Conclusions}

We have developed means to estimate star formation rates accurately
from MIPS 24$\mu$m photometry and other single-band infrared measurements.
The SFR is expressed in Equations 10 and 11. We have converted
these equations into simple fits to expedite applying this result
as a function of observed MIPS 24$\mu$m flux density and redshift. Similar
expressions are derived for future infrared missions, summarized in Equation 14
and Tables 1 and 2. We also determine
an accurate form of the radio-infrared relation for local galaxies
and provide an equivalent expression for the star formation rate in 
terms of 21 cm luminosity. We show that the relation between 
observed 1.4 GHz and 850$\mu$m flux densities in our templates
is not changed at z $\sim$ 2, supporting use of these bands
to determine photometric redshifts for luminous infrared galaxies. 

Our methods apply to infrared-selected galaxies with luminosities
$\ge$ 5 $\times$ 10$^9$ L$_\odot$ and $<$ 2 $\times$ 10$^{12}$ L$_\odot$. 
At lower luminosity, a
significant fraction of the output of newly formed stars escapes
without being absorbed by interstellar dust and re-emitted in the
infrared, while at very high luminosities there are no local examples
of star forming galaxies to guide us in assembling spectral templates. 
Galaxies selected in the ultraviolet will tend to
favor those with lower absorption of the stellar output, although
this appears to be only a moderate effect (Figure 3 of
Iglesias-P\'aramo et al. 2006 or Figure 7 of Buat et al.
2007). Thus, for local galaxies our estimates of star formation
should be generally applicable for luminosities above 3 $\times$ 10$^{10}$ L$_\odot$. 
However, there are indications that the luminosity threshold
for their general use may be higher at high redshift (e.g., Flores et al. 
2004; Buat et al. 2007, Figure 7). 
%We also discuss the effects on our derivations of possible 
%changes in even very luminous infrared galaxy properties at high redshift.
%Using our templates to K-correct 24$mu$m observations at $z\sim 2$
%should not overpredict restframe L(24) by more than a factor of 2 in
%starforming galaxies, with the caveat that AGN contribution to the
%observed 24$mu$m flux can also become more significant as the rest
%bandpass shifts to shorter wavelengths.
Additional work is needed to calibrate the possible systematic
effects at, e.g., z$\sim$ 1-2. 

An intermediate product of this work is a set of accurate spectral energy
distribution templates for infrared galaxies. The construction of these
templates is described in the Appendix and they are available on-line
for other uses. 

\acknowledgments

We are grateful for helpful discussions with 
Phil Appleton, Chad Engelbracht, and Robert
Kennicutt.  We also thank Daniela Calzetti for providing the Pa$\alpha$
measurements; Jennifer Sierchio for resampling the
LIRG and ULIRG SEDs onto a common wavelength scale; 
Eckhard Sturm for providing the PACS response curves; and
Alistair Glasse for providing MIRI response curves.  We also
thank the FIDEL team, led by Mark Dickinson, for the data presented
in Figures 17 and 18.  
This work is based on observations made with the {\it Spitzer} Space 
Telescope, which is operated by the Jet Propulsion Laboratory, California 
Institute of Technology under contract with NASA. This research
has used the NASA/IPAC Extragalactic Database (NED), which is operated
by the Jet Propulsion Laboratory under contract with NASA. It also
makes use of data products from the Two Micron All Sky Survey,
which is a joint project of the University of Massachusetts
and the Infrared Processing and Analysis Center/California Institute
of Technology, funded by NASA and the NSF. Support for this work 
was provided by NASA through JPL/Caltech contract 1255094 and by 
NAG5-12318 from NASA/Goddard to the University of Arizona. 
Funding for the DEEP2 survey has been provided by NSF grants AST95-09298, 
AST-0071048, AST-0071198, AST-0507428, and AST-0507483 as well as NASA LTSA grant NNG04GC89G. 
P.~G. P\'erez-Gonz\'alez acknowledges support from the
grants PNAYA 2006--02358 and PNAYA 2006--15698--C02--02 and from the Ram\'on
y Cajal Program, financed by the Spanish Government and the European Union.

%\bibliography{general}

%\bibliographystyle{apj}

%\appendix
\section{Appendix: Assembly of SED Templates}

Our assembly of SED templates begins with a detailed consideration of eleven
local LIRGs and ULIRGs. We discuss the input data for them and then
how the data were combined consistently into individual templates. 
We next describe how we used overall color trends and these
eleven templates to build average templates for local LIRGs and
ULIRGs that should be representative of the entire population of
such galaxies. Finally, we discuss the derivation of a compatible
set of templates for lower luminosity infrared galaxies.

\subsection{Input Data and Models for Individual LIRGs and ULIRGs}

\subsubsection{Mid-Infrared Data}

The number of ULIRGs that have both high quality data across the electromagnetic spectrum and
are firmly established to be dominated by star formation is small. Our sample is
IRAS 12112+0305, IRAS 14348-1447, IRAS 22491-1808, and Arp 220, all of which were 
studied by Armus et al. (2007) and thus
with high quality IRS spectra, plus IRAS 17208-0018 where a usable spectrum is
available from ISO (Rigopoulou et al. 1999). There are few if any additional
ULIRGs that meet our criteria for both complete data sets and the absence
of AGN. We selected LIRGs from a sample originally selected for Pa$\alpha$ imaging
with NICMOS (and hence over a restricted range of redshift) and to represent
a full range of SED properties. They are NGC 1614, NGC 2369, NGC 3256, 
ESO0320-g030, and Zw049.052 (Alonso-Herrero et al. 2006), plus NGC 4194.   

Our templates for LIRGs make use of previously unpublished spectroscopy obtained with the
Infrared Spectrograph (IRS; Houck et al. 2004) on {\it Spitzer}. The observations (PID
30577) were taken in low resolution, and in mapping mode to cover all or nearly all of
the infrared emitting region of the galaxy. They were
reduced using Cubism (Smith et al. 2007), a package written specifically for optimum 
reduction of IRS mapping observations. The resulting spectral maps were then
combined into a single spectrum to represent the integrated infrared output
of the galaxy. These spectra were supplemented by those of NGC 1614 and NGC 4194
obtained from Brandl et al. (2006). In these cases, the single slit spectra
also include virtually all of the infrared-bright region (after the rescaling
of the short wavelength spectra as described in Brandl et al. (2006)). 
For the ULIRG templates, we made use of published spectra
(Rigopoulou et al. 1999; Armus et al. 2007; Armus, private communication, 2007). 
The IRS slits should include most of the flux from all of these galaxies.
In all cases, the agreement of the spectra with photometry of the full
galaxy verified that we had not missed any significant signal with the spectra.

In addition, we obtained IRAC (Fazio et al. 2004) photometry of the sample galaxies from
the {\it Spitzer} archive (PIDs 32, 108, \& 3672). They were reduced in the SSC
Pipeline Version S14 and 
photometry was obtained within apertures that included the entire galaxies.
A correction for extended emission was applied to the results as recommended
by the {\it Spitzer} Science Center (http://ssc.spitzer, caltech.edu/irac/calib/extcal/). 
We also make use of the IRAC and MIPS photometry
from the SINGS program (Dale et al. 2007) and of IRAS photometry,
where we preferentially use the reductions of Sanders et al. (2003).

\subsubsection{Additional Data and Models}

There are no homogeneous galaxy spectra that
cover the entire 0.4 to 2$\mu$m range. However, as a key input to the templates, we used 
the spectrum from 0.8 to 2.4$\mu$m of the lightly obscured LIRG
NGC 1614 (Alonso-Herrero et al. 2001), obtained with a spectrograph that maintains the relative normalization
of the spectral segments  \citep{Rif06}. We compared it with the GALAXEV (version 2003) model from
Bruzual and Charlot (http://www.cida.ve/~bruzual/bc2003) for a solar
metallicity, 20Myr old population. The agreement of the overall 
SED and spectral features was quite close, within the measurement errors
as well as they could be judged by the examination of the spectrum. 
The GALAXEV model was therefore used to represent a full galaxy spectrum. 

For a whole galaxy, there is a question of the relative contribution of
old and young stars; that is, do faint outer and old components add up
to sufficient flux to influence the overall spectrum. For our sample members, we compared the 
ratios of full galaxy 2MASS K photometry to large beam IRAS 25$\mu$m fluxes with
the similar ratio just for the starburst region of M82 to test
this possibility. In only two of our eleven galaxies (i.e., for ESO0320-g030 and
NGC 2369) was the ratio
more than a factor of two higher than that for the nucleus of M82. This test
suggests that an additional cool stellar population may dominate the near infrared flux
for these two galaxies but that the young stars are prominent in the rest. 

In constructing templates, we force the stellar SED 
to fit large aperture photometry, so any issues about the
details of the spectral features will be
accommodated to first order. A possible exception is broad molecular 
bands, of which the most prominent is the 2.3$\mu$m first overtone CO
absorption\footnote{The CO fundamental is not significantly stronger than the
overtone, due to saturation, and it is diluted by emission by dust.}. 
To test the effect of changes in this band strength, we constructed a template with it reduced
by $\sim$ 0.05 to be representative of an old or low-metallicity stellar population
and tested it on IRAC color-color plots (Donley et al. 2008, Appendix B). It had only
a small effect on the color locus. Some of the less prominent spectral
features may not be accurately represented in our templates, but it is not possible with
current data to make a well-constrained correction for this behavior.
In addition, there are alternative theoretical templates to GALAXEV,
but again the differences are not large enough to be significant
for the templates once the behavior has been forced to fit the photometry.
These statements apply to use of the templates at photometric resolution,
of course. Relatively narrow spectral features may still be inaccurately
represented (e.g., Riffel et al. 2008).

We need to extend the photospheric SED to beyond 5$\mu$m, to merge with the
IRS spectra. There are few galaxy spectra to use as a guide, and even spectra
of stars suitable for population models of galaxies are not common. Fortunately, the photospheric 
spectral behavior of galaxies from 2 to 5$\mu$m should be relatively simple, being
dominated by the coolest giant and supergiant stars present in large numbers.
We therefore used the airborne spectrum of $\beta$ Peg (M2.5 II-III) \citep{Str79}, normalized
over the 2 - 2.3$\mu$m region to the GALEXEV model, for the extension. The
CO absorption strength in this star is similar to that observed in luminous
star forming galaxies, so it provides an excellent surrogate for a true
galaxy spectrum. In this way, we have constructed a consistent stellar 
photospheric continuum appropriate
for a LIRG or ULIRG.  

Between 3 and 5$\mu$m, there is additional emission from warm dust and aromatic
features, (e.g., Lu et al. (2003)). However, there is no significant body of spectroscopy
with a large enough beam to match to the IRS spectra across this region. 
Empirical template libraries are often quite approximate over this range 
(as in Dale \& Helou 2002)). More realistic templates in this region
are important for comparison, for example, with IRAC-color-based methods
for identifying AGN. We therefore determined a typical excess spectrum starting from the ISO
spectrum of M82 reported by St\"urm et al. (2000). We fitted our stellar photospheric
SED to M82 in the near infrared and then subtracted it from the published spectrum
to derive one of the mid-infrared excess alone. 

We obtained additional far infrared, submm, and radio data 
for the sample galaxies from
the NASA Extragalactic Database (NED). Full-galaxy JHK photometry was taken
from the 2MASS extended source database. Full-galaxy optical photometry was
obtained from NED or other sources in the literature (e.g., Surace \& Sanders
2000, Kim et al. 2001, Taylor et al. 2005).

\subsubsection{Template Construction}

To construct a template, we started with our 0.4 - 5$\mu$m stellar photospheric
template and the large beam photometry in the optical and near-IR. 
We used a simple reddening law to match the stellar SED
to the photometry from 0.4 to 2.5$\mu$m. We took the underlying reddening as in
Rieke \& Lebofsky (1985) and assumed that the dust and stars were mixed
in an optically thick configuration. For a few LIRGs, this model
failed in the blue until we added a small amount of reddening in
a foreground screen. We then compared the predicted stellar photospheric
output from 3 to 6$\mu$m with the IRAC full-galaxy photometry. We added
the M82-based infrared-excess SED multiplied by a power law slope
adjustment to the stellar prediction to fit the IRAC photometry. 

The resulting template out to 6$\mu$m always joined in a consistent manner onto
the IRS or ISO spectrum. We used the IRS spectrum to define the template to
past 35$\mu$m. For IRAS 17208-0018, where the useful ISO spectrum stops at 10.7$\mu$m,
we extended to 35$\mu$m using the SED of IRAS 2249 which is very similar
in the region of overlap. The spectra out to 35$\mu$m were interpolated
onto a uniform wavelength sampling (to remove the effects of redshift) using
a Hermite spline technique. 

To determine a far infrared template, we found that a single black body
with wavelength-dependent emissivity provided an adequate fit to all of
the far infrared and submm photometry of each galaxy. In comparison, 
the templates of Dale \& Helou (2002) often gave a poorer fit, with
a spectral peak that was too broad for the SEDs of the high luminosity
galaxies. This behavior is perhaps not surprising given that the Dale \&
Helou templates were developed to fit galaxies of starburst luminosity,
where a broad range of conditions and dust temperatures can contribute
significantly to the overall SED. We optimized modified blackbody fits
to each galaxy, with a resulting range in the fitted temperature of 38 - 64K
and in the coefficient of emissivity of $0.7 < \beta < 1$, where the emissivity
goes as $\lambda^{-\beta}$. Despite the range of parameters, the SEDs
are extremely similar in the submm, with some variation in the location of
the peak of the SED near 100$\mu$m. In some cases, we found that this simple
fit failed to join smoothly onto the IRS spectrum near 35$\mu$m and that the
spectrum beyond this wavelength was very noisy. We solved
this problem by interpolating linearly from the last high-weight
IRS points (near 35$\mu$m) to 63$\mu$m. 

For the radio, we again gathered all available data, primarily from NED and from
Condon et al. (1991). Our fits to these data assumed an intrinsic power law
slope of -0.7 at high frequencies. The slightly steeper slope characteristic
of starburst luminosity galaxies (-0.8) gave slightly poorer fits. 
As necessary, the power law index was
changed to a (single) lower value to fit the low frequency data. 
The final templates are shown in Figures 1, 2, 3, and 4 along with the photometry
used to constrain them. It can be seen that the templates provide a
good representation of the available photometry, based on a plausible
detailed SED. They are provided in Table 3.

We also show a Dale and Helou (2002; Dale, web site http://physics.uwyo.edu/$\sim$ddale/)
template with $\alpha = 1.5$, appropriate for a highly active galaxy 
($\alpha$ is defined in Dale \& Helou (2002)) . 
The Dale and Helou (2002) templates were developed to fit
the SEDs of moderate luminosity star-bursting galaxies (L $<$ 10$^{11}$ L$_\odot$).
Although they are often used as templates for high-redshift LIRGs and
ULIRGs, Figures 3 and 4 show that they differ significantly from the
observed behavior of local ULIRGs in the mid and far infrared 
and hence are probably not ideal for representing higher-redshift ones. 
Short of about 5$\mu$m, the Dale and Helou templates are relatively
schematic and do not include accurate stellar photospheric data. 
In the 10$\mu$m region, they omit silicate absorption, which is
not strong in the starburst luminosity range for which they were
optimized. However, with increasing luminosity and the accompanying
increasing extinction, this feature can become quite strong in, 
LIRGs and ULIRGs. In addition, at high infrared luminosities,
our far infrared SEDs are more peaked. In large part this results
from our use of IRS spectra for the 15 to 35$\mu$m range, whereas
Dale and Helou (2002) had to fill it in largely by interpolation.

In addition, we show a Chary \& Elbaz (2001) template for L(TIR) = $2 \times 10^{12}$ L$_\odot$.
It also deviates significantly from our templates.
The CE templates have strongly suppressed aromatic features at high
luminosities (not consistent with the ISO (Rigopouplou et al. 1999) or 
IRS spectra (e.g., Armus et al. 2007, this work)). Their behavior in
the far infrared is similar to that of the Dale and Helou (2002) models.
Finally, the models of Siebenmmorgen \& Kr\"ugel (2007) appear to have more
cold dust than our templates at low luminosities and weak silicate
absorption at high luminosity. That is, the availability of new
data, particularly the long wavelength IRS spectra, has resulted in
our templates differing significantly from those produced on the
basis of ISO and IRAS data alone.

\subsection{Average LIRG and ULIRG Templates}

\subsubsection{Behavior of Infrared Galaxy Colors}

Although the templates for individual galaxies are useful, for example
to judge probable ranges of spectral behavior (e.g., Donley et al. 2008),
for many applications we need average templates. To begin the derivation
of such averages, we consider the behavior of the mid- and far-infrared
colors of LIRGs and ULIRGs.
.

%\subsubsubsection{Relation of Individual Bands to TIR}

\medskip
\centerline{6.2.1.1. \it Relation of Individual Bands to TIR}
\medskip

First, we display the performance of the 8 (IRAC), 24 (MIPS), and 12 and 60$\mu$m (IRAS) 
bands as predictors of L(TIR). We define the luminosity (in L$_\odot$) at a given
wavelength to be proportional to $\nu f_\nu$ and compute L(TIR)
as described in Sanders et al. (2003; see also Sanders \& Mirabel 1996). They assumed a 
cosmology with H$_o$ = 75 km s$^{-1}$ Mpc$^{-1}$, $\Omega_M = 0.3$, and $\Omega_\Lambda = 0.7$. 
They define:

\begin{equation}
\label{eq20}
L(TIR)=4\pi D_L^2 \;(1.8\times 10^{-14}[13.48f_{12} +5.16f_{25} 
+2.58 f_{60} +f_{100} ]),
\end{equation}

\noindent
where the fluxes are for the IRAS bands.
Our 8$\mu$m photometry for
starburst galaxies is from Dale et al. (2007) and Engelbracht et al. (2008), complemented for
LIRGs and ULIRGs by data taken from the {\it Spitzer} archive. For
the IRAS data, we have used the bright galaxy sample reductions
whenever possible (Sanders et al. 2003). 

Figure 13 compares L(8) with L(TIR). A linear fit for the roughly solar
metallicity galaxies and 8.5 $\le$ log(L(TIR)) is

\begin{equation}
\label{eq21}
\log (L(TIR))=(-1.008\pm 0.457)+(1.190\pm 0.047)\log (L(8))
\end{equation}

\noindent
The scatter around this relation is large, with an rms of 0.44 dex.
The curve appears to be somewhat nonlinear, with a significantly
flatter slope at ULIRG luminosities (a behavior we term saturation).  
If we only fit the data for 
L(TIR) $>$ 10$^{10}$ L$_\odot$, we obtain

\begin{equation}
\label{eq22}
\log (L(TIR))=(-1.46\pm 1.12)+(1.24\pm 0.11)\log (L(8))
\end{equation}

\noindent
with still a large scatter of 0.35 dex.

Figure 14 compares L(12) with L(TIR). After rejecting one 
outlier with very faint 12$\mu$m emission (NGC 1316), a
fit to the results for 8.5 $\le$ log(L(TIR)) yields

\begin{equation}
\label{eq23}
\log (L(TIR))=(-0.947\pm 0.324)+(1.197\pm 0.034)\log (L(12))
\end{equation}

\noindent
with a scatter of 0.21 dex. In this case, the relation
appears to be slightly non-linear in log-log space in the sense that
the most luminous galaxies are underluminous at 12$\mu$m (a mild
case of saturation). 
A fit for log(L(TIR) $\ge$ 10 is preferable for high luminosities:

\begin{equation}
\label{eq24}
\log (L(TIR))=(-1.94\pm 0.71)+(1.296\pm 0.070)\log (L(12))
\end{equation}

Figure 15 compares L(24) and L(TIR). We converted the IRAS
measurements to equivalent ones for MIPS at 24$\mu$m through the average
ratio for well-measured galaxies, f(IRAS)/f(MIPS) = 1.16 $\pm$ 0.02. 
Our relation applies to the IRAS and MIPS photometry with no bandpass 
or other corrections applied. A fit 
to the data for 8.5 $\le$ log(L(TIR)) is:

\begin{equation}
\label{eq25}
\log (L(TIR))=(1.445\pm 0.155)+(0.945\pm 0.016)\log (L(24))
\end{equation}

\noindent
with a scatter of 0.13 dex. Neither the fit nor the scatter
is significantly different for a linear fit only for log (L(TIR)) $\ge$ 10. 

Figure 16 compares L(60) and L(TIR). 
A fit for 8.5 $\le$ log(L(TIR)) is

\begin{equation}
\label{eq26}
\log (L(TIR))=(1.183\pm 0.101)+(0.920\pm 0.010)\log (L(60))
\end{equation}

\noindent
The fit and the scatter (0.08 dex) around it are virtually the same if the range is
restricted to log(L(TIR)) $>$ 10. 

These fits indicate that the
common approach of assuming that the L(TIR) is linearly proportional
to the luminosity in some mid-infrared band can be improved by taking
account of the small deviations from proportionality. The use of
SED templates to carry out this conversion (e.g., Le Floc'h et al.
2005; Marillac et al. 2006) in principle solves this problem,
but only by placing strong reliance on the accuracy of the templates. 

The linear fits
show that the performance of the 24$\mu$m band in predicting L(TIR)
is only slightly worse than that at 60$\mu$m, perhaps a surprising
result since 60$\mu$m is generally the dominant band in determining
L(TIR). We find that the 12$\mu$m band remains useful in predicting
L(TIR) but now with substantial scatter. That is, the 24$\mu$m MIPS
band can be used as a reasonably accurate measure of L(TIR)
up to redshifts of z $\sim$ 1. The behavior at 8$\mu$m
is interesting. Up to 10$^{11}$ L$_\odot$, it behaves reasonably
well, in agreement with previous work (e.g., Roussel et al. 2001). However,
it is not safe to conclude from this behavior that it is equally
useful as a L(TIR) measure above this luminosity, where the
saturation effect becomes strong and the scatter is large. At
redshifts of z $\sim$ 2, the 24$\mu$m MIPS band is at a rest
wavelength of $\sim$ 8$\mu$m and is widely used as a measure
of L(TIR). Fortunately, it appears that the saturation phenomenon
may be much weaker at this redshift (Rigby et al. 2008), so with 
a recalibration valid results may still be obtained.

%Yun et al. (2001) compared observed radio and infrared flux
%densities without regard to corrections connected with redshift 
%and found a constant ratio of radio to infrared luminosity.
%For the great majority of their sample, this approach was 
%appropriate, because the high-weight IRAS detections do not
%extend to large distances for starburst luminosity galaxies. However,
%for the most luminous galaxies in their sample, redshifts of
%z $\sim$ 0.05 are typical. Even at this modest redshift, the very
%steep drop in infrared SED toward short wavelengths from 60$\mu$m
%can result in significant K-corrections in the infrared luminosity.

%We show the effect of these corrections in Figure 9, where we plot the ratio 
%q = log(f(60$\mu$m)/f(1.4GHz)) vs. log(L(60)) for the galaxies in the Yun
%et al. study. We assumed a slope of -0.7 for the radio
%spectrum and one of -2.2 for the local slope of the infrared
%SED. The resulting corrections in q produce a noticeable increase
%with growing L(60), which can be fitted with

%\begin{equation}
%\label{eq8}
%q=(-0.014\pm 0.857)+(0.213\pm 0.075)L(60)
%\end{equation}

%\noindent
%Although the scatter is large, the slope is significant at nearly
%the 3-$\sigma$ level. Therefore, in the following we take
%the radio/IR relation to be constant (for 1.4GHz and 60$\mu$m) 
%at q = 2.33 for L$<$10$^{11}$L$_\odot$
%and to be given by the above expression for larger luminosities.
%This relation provides an alternative means to estimate the
%L(TIR) and the SFR in the absence of 24$\mu$m or far infrared data.

\subsubsection{Average Templates}
The challenge in deriving average templates from our eleven individual ones is
to relate them rigorously to the behavior of a much larger sample, for which
many members do not have all the data required to constrain templates
in detail. We approached this challenge by building a spreadsheet that
allows us to combine (averaging in the logarithm) the eleven templates
with different weights. The spreadsheet conducts synthetic photometry
on the combined SED, which we compare with the average relations represented
by the fits to the colors derived in the preceding section. To minimize the effects of
nonlinearity in the relations at 8 and 12$\mu$m, all the fits to the colors 
were for L(TIR) $\ge$ 10$^{10}$ L$_\odot$. In addition, we combine  
SEDs for galaxies of roughly the appropriate luminosity (e.g., no ULIRG
components in the fit for a low luminosity LIRG template). We use
synthetic colors in 25/8$\mu$m, 25/12$\mu$m, and 60/25$\mu$m as simultaneous
constraints to select an appropriate combination of the individual templates. 
We found that any combination of individual SEDs that fitted
these photometric colors yielded virtually identical average SEDs. To the
extent there were minor differences, we selected the fit that also
provided a smooth progression of spectral behavior from one luminosity
to another.  

\subsection{Templates at Intermediate Luminosity (between $5 \times 10^9$ and $10^{11} L_\odot$)}

Much of the preparation for constructing similar average templates for intermediate-luminosity
infrared galaxies has been completed by Dale et al. (2007) and Smith et al. (2007).
The first of these papers fits Dale \& Helou (2002) templates to a
large number of well-studied galaxies, determining the value for
the $\alpha$ parameter that controls the far infrared shape
of the template for each galaxy. Since this work provides fits
to a large number of individual galaxies, it is analogous to the fits to
individual LIRGs and ULIRGs in this paper and is similarly useful to examine the
range of behavior. The second paper constructs
"noise-free" templates in the mid-infrared from IRS spectra of
a similar sample of galaxies. To construct average templates in the
same style as those we have built for the high luminosity galaxies
requires that the Dale \& Helou template fitting and the Smith et al.
spectral templates be mapped to L(TIR), and then that the two template
sets be joined consistently near the long wavelength limit of
the Smith et al. spectral templates.  

To determine the far infrared behavior of the Dale \& Helou templates,
we took all the galaxies in their sample that are also in
the IRAS bright galaxy sample (BGS; Sanders et al. 2003) and correlated
L(TIR) with $\alpha$. To extend this correlation to high
luminosities, we also fitted Dale \& Helou templates to the
eleven galaxies for which we have determined templates. The
fit is

\begin{equation}
\label{eq27}
\alpha =10.096-0.741\;\log (L(TIR))
\end{equation}

\noindent 
from log(L(TIR)) = 9.5 to 11.6 and $\alpha$ = 1.5 
at higher luminosities. We used this fit to select
the appropriate model SED up through log(L(TIR)) = 11. 

To map the Smith et al. (2007) templates onto L(TIR)
we again used the IRAS BGS (Sanders et al. 2003), this
time to determine the relationship between the ratio of flux densities
at 12 and 25$\mu$m and L(TIR). We fitted with gaussians the distributions
of the 25/12$\mu$m flux density ratios over sliding
L(TIR) bins 0.5 dex wide. Fitting rather than taking
a straight average provides immunity to a small number
of extreme ratios that might otherwise significantly bias
the result. We then combined the four noiseless templates
of Smith et al. with various weights 
and did synthetic photometry on the resulting
average SED, adjusting the combination to match the behavior
of the galaxies in the BGS. We found that the resulting average
template had very little dependence on the input templates,
so long as it matched the photometry. We also remark that this
approach will not work well at log(L(TIR))$<$ 9.5 or $>$ 11,
because the quality and sample size for the IRAS photometry becomes
too small. 

Finally, to join the Dale \& Helou far infrared templates and
the Smith et al. spectral ones, we adopted a relative
normalization that matched the 70 and 160$\mu$m photometry
provided with the templates by Smith et al. {\it and} also
matched synthetic colors for the ratio of IRAS 60 and 25$\mu$m
flux densities at the appropriate L(TIR). Because the templates
did not join smoothly, we interpolated the SEDs linearly between 36.5
and 70.4$\mu$m. 

We show the full set of templates in Figures 5 and 6. They are
tabulated in Table 4.  The values of $f_\nu$ in Table 4 have been normalized
to the correct L(24) for a galaxy at a distance of 10 Mpc.
These templates may under-represent galaxies
dominated by very cold dust because of the subtle biases in
sample selection from {\it IRAS} data; submm measurements
show that there are a reasonably large number of such galaxies
(C. Willmer, private communication, 2008). 

\subsection{Template Behavior at High Redshift} 

So far, we have only considered the SEDs of local galaxies.
We now compare our template set with measurements of galaxies
at redshifts up to z $\sim$ 2.5. 

\subsubsection{Behavior of IRAC Colors}

Lacy et al. (2004) and Stern et al. (2005) proposed two
forms of IRAC color-color diagram to separate star forming
galaxies from those with substantial AGN contributions to
their mid-infrared emission. The reader is referred to 
Donley et al. (2008) for a demonstration of the use of
our templates to test and refine these arguments. 
This paper shows the locus on IRAC color-color diagrams 
expected for star-formation-dominated galaxies represented by our templates.
Above z = 2.3, the 6.2$\mu$m aromatic band has moved
out of the MIPS 24$\mu$m band and extremely luminous
star formation would be required for a detection. Instead,
it is likely that the detections in this range of
redshift are boosted by AGN. Below this redshift, the great majority
of the 24$\mu$m-detected galaxies have IRAC colors that
place them within the contours determined by redshifting
our templates through the IRAC bands. Scientifically,
this behavior indicates that the IRAC band signals for the
majority of 24$\mu$m-detected galaxies are dominated
by stellar processes. For our templates, it indicates
that there are no significant systematic errors in their representation
of the stellar-driven SEDs as detected in the IRAC bands 
over the entire relevant redshift range. 

\subsubsection{Aromatic Band Strength}

Figure 17 compares our templates with the observed 24-to-70$\mu$m
colors of galaxies as a function of redshift 
using Spitzer/MIPS fluxes from the FIDEL legacy survey,
matched with DEEP2 spectroscopic redshifts for 
galaxies in the Extended Groth Strip (Dickinson et al. 2007; Davis et al. 2007).
Out to z $\sim$ 1,
the templates nicely envelope the observed colors, showing that
they represent the full range of spectral behavior. In fact, 
the figure indicates that Zw049.052 may have extreme SED
behavior that is uncommon. However, above
z $\sim$ 1, galaxies with ratios of flux density at 70 and
24$\mu$m as large as predicted by the most extreme local
templates are very rare, even though our method for looking
for such galaxies is strongly biased toward their discovery
(since it requires a detection at 70$\mu$m; in addition, 
blending in the large beam at 70$\mu$m may result in
overestimation of the flux density from a source). 

Figure 18 is another illustration of the same point. It shows 
the photometry for FIDEL/EGS galaxies with photometric redshifts of $\sim$ 0.7,
along with our average templates, all normalized at 24$\mu$m. 
The spread in 70$\mu$m points is matched well by the spread
in behavior of the templates. There is a very large spread in
the points in the stellar photospheric region, which probably
reflects the variation in the contribution of the old
stellar populations in the individual galaxies. Similar
plots that isolate galaxies with higher observed 24$\mu$m
luminosity show a tendency for their 70-24$\mu$m colors to cluster near
the "blue" zone.  This behavior is illustrated by the bar showing  
the distribution of values to the right of the 70$\mu$m points.
Interpreting this result requires determination
of the extent to which this sample is contaminated by AGN, 
which we reserve for a future publication. 

The aromatic bands for very high luminosity galaxies at z $\sim$ 2 
behave similarly to those for lower-luminosity galaxies
locally - that is, they are stronger and account for a larger 
fraction of the bolometric infrared output of the galaxies than is
the case for the more extreme local ULIRGs (e.g., Sajina et al. 2007; 
Papovich et al. 2007, Pope et al. 2008, Rigby et al. 2008; Farrah et al. 2008). 
The absence of galaxies with large ratios of
70 to 24$\mu$m fluxes in Figure 17 provides additional evidence
for this overall pattern of SED behavior.

Two possibilities have been suggested to explain
this result (Rigby et al. 2008, Farrah et al. 2008). 
First, the high redshift ULIRGs may be undergoing
extended star formation rather than the extreme nuclear events that occur
locally. This possibility is consistent with interferometric measurements, 
(e.g. Tacconi et al. 2006, Younger et al. 2008) indicating that, although compact,
the high redshift ULIRGs are significantly more extended than local ones
(e.g., Biggs \& Ivison 2008, Sakamoto et al. 2008). 
The extended star forming regions are likely to
have lower optical depths than the local, nuclear ones. 
The high duty cycle of star formation in these
objects (Daddi et al. 2007) would point in the same direction.
Second, the behavior might be the result of lower metallicity,
which would reduce the dust content of the star forming
regions and reduce their optical depths in that way. In
fact, Engelbracht et al. (2008) have seen a similar trend
of aromatic feature behavior in local galaxies 
with reduced metallicity down to about 1/3 solar
 
Whatever the cause, given the widespread use of the
MIPS 24$\mu$m photometry as a star formation indicator
at high redshift, it is important to allow for this
shift in behavior. Rigby et al. (2008) and our Figure 17 demonstrate
that the behavior can equivalently be described in
terms of a reduction in the proportion of galaxieswith small ratios of L(8)/L(TIR), the low
lying outliers at high luminosity in Figure 13. 
Consequently, it is likely that the scatter in this
ratio is smaller at z $\sim$ 2 than locally, making rest 8$\mu$m
photometry easier to interpret as an indicator of
L(TIR) and star formation rates (once it has been
adequately calibrated at this redshift). 
The average templates presented
here should be appropriate to first order because they are fitted
to the trend of colors from 10$^{10}$ L$_\odot$ upward, and hence the
fits are strongly influenced by the lower-luminosity local galaxies
whose behavior appears to match that of the high-luminosity ones
at high redshift. At intermediate redshifts, the templates may
therefore slightly under-represent the influence of the extreme
SED examples (e.g., Arp 220). 

To evaluate the possible uncertainties near z = 2 we have fitted the trend
of L(8) vs L(TIR) in Rigby et al. (2008) for the stacked
SEDs of Daddi et al. (2005) and Papovich et al. (2007). 
These values probably give the most representative measure
of the SED behavior at this redshift.
We have assumed that the 12 and 25$\mu$m colors are unchanged from
the local behavior and have de-emphasized 60$\mu$m in the fits
(all wavelengths apply to the rest SED). We have used these templates
to compare with the ones at the same L(TIR) fitted to local galaxies,
and converted the results into rest 24$\mu$m luminosities. 
For log(L(TIR)) = 12 and 12.5, we estimate offsets of factors of about 1.2
and 1.3, respectively, 
in the sense that using our local templates will result in an overestimate of
the intrinsic 24$\mu$m flux density by these amounts; larger offsets 
are likely at log(L(TIR)) = 13. 
These estimates are very approximate, since the SEDs of such luminous
purely star-forming galaxies are not well understood either 
locally or at z $\sim$ 2.  In fact, there are significant differences 
among the sets of templates proposed for such high luminosities, with 
no local analogs to constrain them. Further work is needed to resolve 
these discrepancies and to calibrate the use of 24$\mu$m data to estimate 
star formation rates in these cases.

\clearpage

\begin{deluxetable}{rrrrrrr}
\tabletypesize{\scriptsize}
\tablecaption{SFR(flux) fit coefficients A and B as a function of redshift for MIPS and PACS
\tablenotemark{a}\label{fitpar_table}}
\tablewidth{0pt}
\tablehead{
%\colhead{z} & \colhead{A_{24}(z)} &  \colhead{B_{24}(z)} & \colhead{A_{24}(z)} &  \colhead{B_{24}(z)}\\
%
$z$ & $A_{24}$ &  $B_{24}$ & $A_{70}$ & $B_{70}$ & $A_{100}$ & $B_{100}$ \\
}
\startdata
 0.0 &  0.417 &  1.032 & -0.591 &  0.964 & -1.046 &  1.131 \\ 
 0.2 &  0.502 &  1.169 & -0.498 &  0.925 & -0.886 &  1.034 \\ 
 0.4 &  0.528 &  1.272 & -0.444 &  0.913 & -0.752 &  0.970 \\ 
 0.6 &  0.573 &  1.270 & -0.391 &  0.913 & -0.670 &  0.935 \\ 
 0.8 &  0.445 &  1.381 & -0.350 &  0.921 & -0.620 &  0.920 \\ 
 1.0 &  0.358 &  1.565 & -0.305 &  0.931 & -0.581 &  0.913 \\ 
 1.2 &  0.505 &  1.745 & -0.252 &  0.942 & -0.544 &  0.912 \\ 
 1.4 &  0.623 &  1.845 & -0.195 &  0.954 & -0.516 &  0.916 \\ 
 1.6 &  0.391 &  1.716 & -0.142 &  0.971 & -0.487 &  0.922 \\ 
 1.8 &  0.072 &  1.642 & -0.105 &  0.997 & -0.455 &  0.929 \\ 
 2.0 &  0.013 &  1.639 & -0.076 &  1.028 & -0.418 &  0.936 \\ 
 2.2 &  0.029 &  1.646 & -0.056 &  1.067 & -0.383 &  0.943 \\ 
 2.4 &  0.053 &  1.684 & -0.054 &  1.114 & -0.346 &  0.954 \\ 
 2.6 &  0.162 &  1.738 & -0.067 &  1.164 & -0.311 &  0.965 \\ 
 2.8 &  0.281 &  1.768 & -0.079 &  1.210 & -0.279 &  0.979 \\ 
 3.0 &  0.371 &  1.782 & -0.075 &  1.241 & -0.256 &  0.998 \\

\enddata
\tablenotetext{a}{$A(z)$ and $B(z)$ are to be used in Equation 14 as the 
intercept and slope of the relation of log SFR on observed IR flux.}

\end{deluxetable}

%\clearpage
\bigskip

\begin{deluxetable}{rrrrrrr}
\tabletypesize{\scriptsize}
\tablecaption{SFR(flux) fit coefficients A and B as a function of redshift for MIRI
\tablenotemark{a}\label{fitpar_table2}}
\tablewidth{0pt}
\tablehead{
%\colhead{z} & \colhead{A_{24}(z)} &  \colhead{B_{24}(z)} & \colhead{A_{24}(z)} &  \colhead{B_{24}(z)}\\
%
$z$ & $A_{18}$ &  $B_{18}$ & $A_{21}$ & $B_{21}$ & $A_{25}$ & $B_{25}$ \\
}
\startdata
 0.0 &  0.634 &  1.278 &  0.547 &  1.114 &  0.346 &  1.002 \\ 
 0.2 &  0.756 &  1.243 &  0.566 &  1.284 &  0.462 &  1.110 \\ 
 0.4 &  0.566 &  1.391 &  0.652 &  1.260 &  0.446 &  1.273 \\ 
 0.6 &  0.542 &  1.714 &  0.512 &  1.373 &  0.534 &  1.264 \\ 
 0.8 &  1.067 &  1.936 &  0.408 &  1.602 &  0.510 &  1.283 \\ 
 1.0 &  0.663 &  1.761 &  0.638 &  1.834 &  0.329 &  1.443 \\ 
 1.2 &  0.171 &  1.623 &  0.737 &  1.837 &  0.281 &  1.658 \\ 
 1.4 &  0.100 &  1.623 &  0.337 &  1.674 &  0.555 &  1.835 \\ 
 1.6 &  0.331 &  1.693 &  0.070 &  1.633 &  0.688 &  1.892 \\ 
 1.8 &  0.396 &  1.762 &  0.058 &  1.631 &  0.355 &  1.737 \\ 
 2.0 &  0.473 &  1.801 &  0.132 &  1.675 &  0.026 &  1.633 \\ 
 2.2 &  0.757 &  1.809 &  0.287 &  1.750 & -0.125 &  1.622 \\ 
 2.4 &  1.131 &  1.645 &  0.379 &  1.783 & -0.063 &  1.642 \\ 
 2.6 &  1.517 &  1.395 &  0.476 &  1.791 &  0.059 &  1.677 \\ 
 2.8 &  1.980 &  0.736 &  0.788 &  1.743 &  0.084 &  1.729 \\ 
 3.0 &  2.119 &  0.390 &  1.098 &  1.614 &  0.163 &  1.781 \\

\enddata
\tablenotetext{a}{$A(z)$ and $B(z)$ are to be used in Equation 14 as the 
intercept and slope of the relation of log SFR on observed IR flux.}

\end{deluxetable}

\clearpage

\begin{landscape}

\begin{deluxetable}{cccccccccccc}
\tabletypesize{\scriptsize}
\tablecaption{Individual Galaxy Templates \label{data_table}}
\tablewidth{0pt}
\tablehead{
\colhead{Wavelength} & \colhead{NGC 1614} & \colhead{NGC 2369} & 
\colhead{NGC 3256} & \colhead{NGC 4194} & \colhead{IRAS1211} &
\colhead{IRAS1434} & \colhead{IRAS17208} & \colhead{IRAS2249} & 
\colhead{Arp 220} & \colhead{ESO0320-g030} &
\colhead{Zw049.057} \\
\colhead{($\mu$m)}  & \colhead{(Jy)} & \colhead{(Jy)} & \colhead{(Jy)}
& \colhead{(Jy)} & \colhead{(Jy)} & \colhead{(Jy)} & \colhead{(Jy)} & 
\colhead{(Jy)} & \colhead{(Jy)} & \colhead{(Jy)} & \colhead{(Jy)}
}
\startdata

0.400 &	0.0171 & 0.0199 & 0.0837 & 0.0191 & 0.000681 & 0.00106 & 0.00197 & 0.00117 & 0.0120 & 0.0197 & 0.00264 \\
0.405 &	0.0181 & 0.0213 & 0.0883 & 0.0203 & 0.000720 & 0.00112 & 0.00209 & 0.00123 & 0.0127 & 0.0210 & 0.00280 \\
0.410 &	0.0118 & 0.0141 & 0.0576 & 0.0133 & 0.000470 & 0.00073 & 0.00136 & 0.00080 & 0.0083 & 0.0138 & 0.00183 \\
0.415 &	0.0178 & 0.0215 & 0.0868 & 0.0201 & 0.000710 & 0.00110 & 0.00206 & 0.00121 & 0.0125 & 0.0210 & 0.00276 \\
0.420 &	0.0172 & 0.0210 & 0.0836 & 0.0194 & 0.000685 & 0.00106 & 0.00199 & 0.00116 & 0.0121 & 0.0205 & 0.00266 \\
0.425 &	0.0180 & 0.0223 & 0.0877 & 0.0205 & 0.000720 & 0.00112 & 0.00209 & 0.00121 & 0.0127 & 0.0217 & 0.00280 \\
0.430 &	0.0177 & 0.0220 & 0.0858 & 0.0201 & 0.000705 & 0.00110 & 0.00205 & 0.00119 & 0.0124 & 0.0214 & 0.00274 \\
0.435 &	0.0172 & 0.0217 & 0.0836 & 0.0196 & 0.000689 & 0.00107 & 0.00200 & 0.00115 & 0.0121 & 0.0210 & 0.00268 \\

\enddata

\tablecomments{The complete version of this table is in the electronic edition 
of the Journal.  The printed edition contains only a sample.}

\end{deluxetable}
%\end{landscape}

\clearpage

%\begin{landscape}

\begin{deluxetable}{ccccccccccccccc}
\tabletypesize{\scriptsize}
\tablecaption{Average Templates\label{avg_table}}
\tablewidth{0pt}
\tablehead{
\colhead{Wave} & 
\colhead{lgL$_{TIR}$} &
\colhead{lgL$_{TIR}$} &
\colhead{lgL$_{TIR}$} &
\colhead{lgL$_{TIR}$} &
\colhead{lgL$_{TIR}$} &
\colhead{lgL$_{TIR}$} &
\colhead{lgL$_{TIR}$} &
\colhead{lgL$_{TIR}$} &
\colhead{lgL$_{TIR}$} &
\colhead{lgL$_{TIR}$} &
\colhead{lgL$_{TIR}$} &
\colhead{lgL$_{TIR}$} &
\colhead{lgL$_{TIR}$} &
\colhead{lgL$_{TIR}$} \\
\colhead{} &
\colhead{9.75} & 
\colhead{10.00} & 
\colhead{10.25} & 
\colhead{10.50} & 
\colhead{10.75} & 
\colhead{11.00} &
\colhead{11.25} &
\colhead{11.50} &
\colhead{11.75} &
\colhead{12.00} &
\colhead{12.25} &
\colhead{12.50} &
\colhead{12.75} &
\colhead{13.00} \\  
\colhead{}            &
\colhead{lgL$_{24}$} &
\colhead{lgL$_{24}$} &
\colhead{lgL$_{24}$} &
\colhead{lgL$_{24}$} &
\colhead{lgL$_{24}$} &
\colhead{lgL$_{24}$} &
\colhead{lgL$_{24}$} &
\colhead{lgL$_{24}$} &
\colhead{lgL$_{24}$} &
\colhead{lgL$_{24}$} &
\colhead{lgL$_{24}$} &
\colhead{lgL$_{24}$} &
\colhead{lgL$_{24}$} &
\colhead{lgL$_{24}$} \\
\colhead{} &
\colhead{8.79} & 
\colhead{9.05} & 
\colhead{9.32} & 
\colhead{9.58} & 
\colhead{9.84} & 
\colhead{10.11} &
\colhead{10.37} &
\colhead{10.64} &
\colhead{10.90} &
\colhead{11.17} &
\colhead{11.43} &
\colhead{11.70} &
\colhead{11.96} &
\colhead{12.22} \\  
\colhead{($\mu$m)} &
\colhead{(Jy)} & 
\colhead{(Jy)} & 
\colhead{(Jy)} & 
\colhead{(Jy)} & 
\colhead{(Jy)} & 
\colhead{(Jy)} & 
\colhead{(Jy)} & 
\colhead{(Jy)} & 
\colhead{(Jy)} & 
\colhead{(Jy)} & 
\colhead{(Jy)} & 
\colhead{(Jy)} & 
\colhead{(Jy)} & 
\colhead{(Jy)} \\ 
}
\startdata

 4.02 & ... & ... & ... & ... & ... & ... & 1.3570 & 2.282 & 2.849 & 4.449 &  5.845 &  8.539 & 16.42 & 29.78 \\ 
 4.06 & ... & ... & ... & ... & ... & ... & 1.3468 & 2.272 & 2.849 & 4.483 &  5.931 &  8.704 & 16.57 & 29.78 \\ 
 4.10 & ... & ... & ... & ... & ... & ... & 2.0877 & 3.593 & 4.827 & 8.158 & 11.284 & 17.284 & 28.64 & 48.53 \\ 
 4.14 & ... & ... & ... & ... & ... & ... & 1.6360 & 2.800 & 3.672 & 6.055 &  8.243 & 12.399 & 21.65 & 37.65 \\ 
 4.18 & ... & ... & ... & ... & ... & ... & 1.4905 & 2.548 & 3.315 & 5.443 &  7.366 & 11.003 & 19.59 & 34.32 \\ 
 4.22 & ... & ... & ... & ... & ... & ... & 1.3451 & 2.296 & 2.964 & 4.819 &  6.509 &  9.648 & 17.52 & 30.99 \\ 
 4.26 & ... & ... & ... & ... & ... & ... & 1.3382 & 2.289 & 2.982 & 4.888 &  6.616 &  9.894 & 17.75 & 31.15 \\ 

\enddata

\tablecomments{The templates for log(L(TIR)) $> 12.25$ are based on 
extrapolation and are likely to have reduced accuracy. 
The complete version of this table is in the electronic edition 
of the Journal.  The printed edition contains only a sample.}

\end{deluxetable}

\clearpage
\end{landscape}

%\clearpage

\begin{figure}
\epsscale{1.0}
%\plotone{templates1}
\plotone{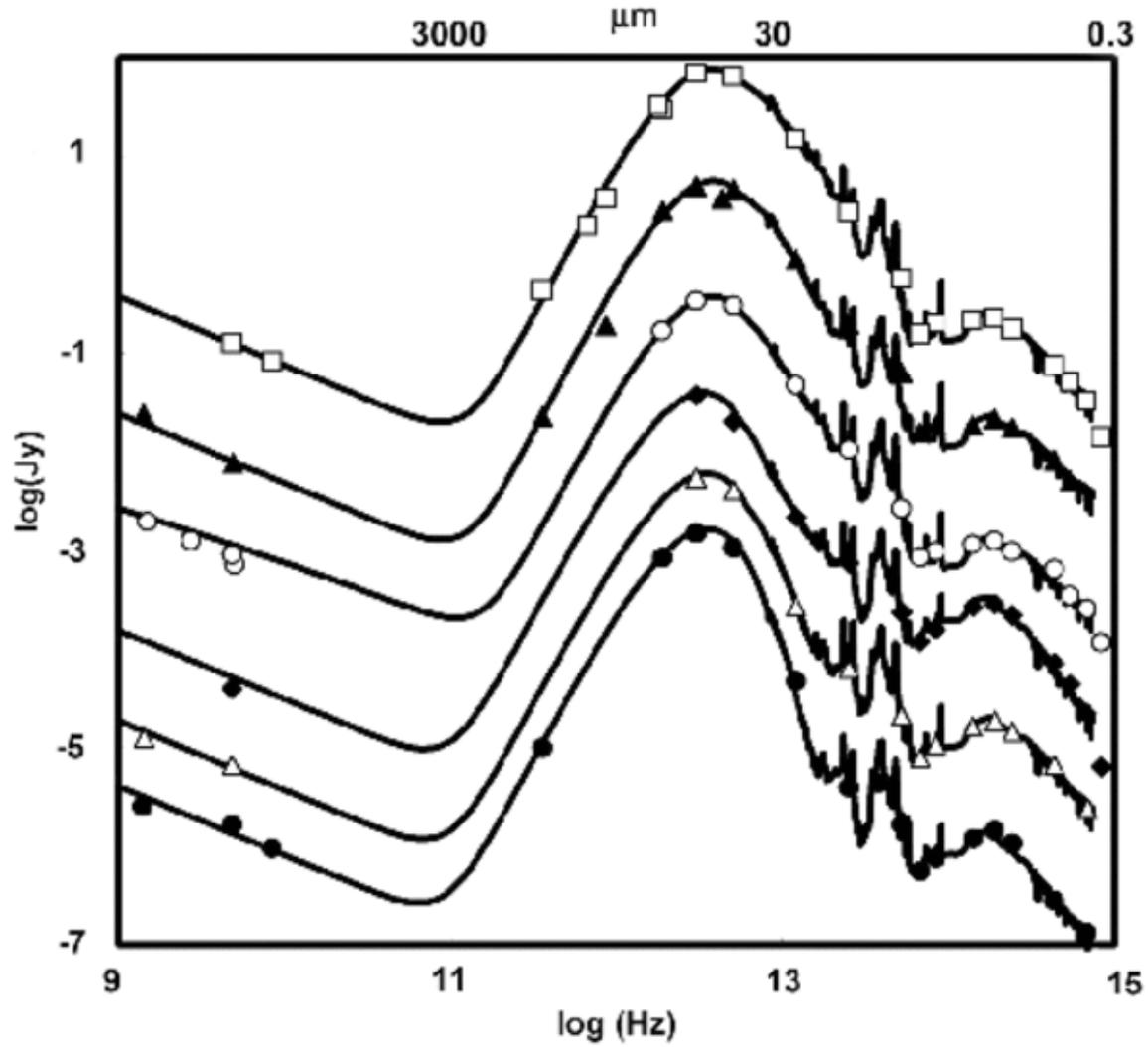}
\caption{Full LIRG SEDs and defining photometric points. Between 5 and 36$\mu$m the SEDs are based on IRS spectra rather than the photometry. The curves have been offset for clarity; from top to bottom the galaxies and offset factors are: 1.) NGC 1614 (2); 2.) NGC 4194 (0.2); 3.) 
NGC 3256 (0.003); 4.) NGC 2369 (0.001); 5.) ESO 0320-g030 (0.0001); and 
6.) Zw 049.057 (0.00005).  
\label{01}}
\end{figure}

\clearpage

\begin{figure}
\epsscale{1.0}
%\plotone{templates2}
\plotone{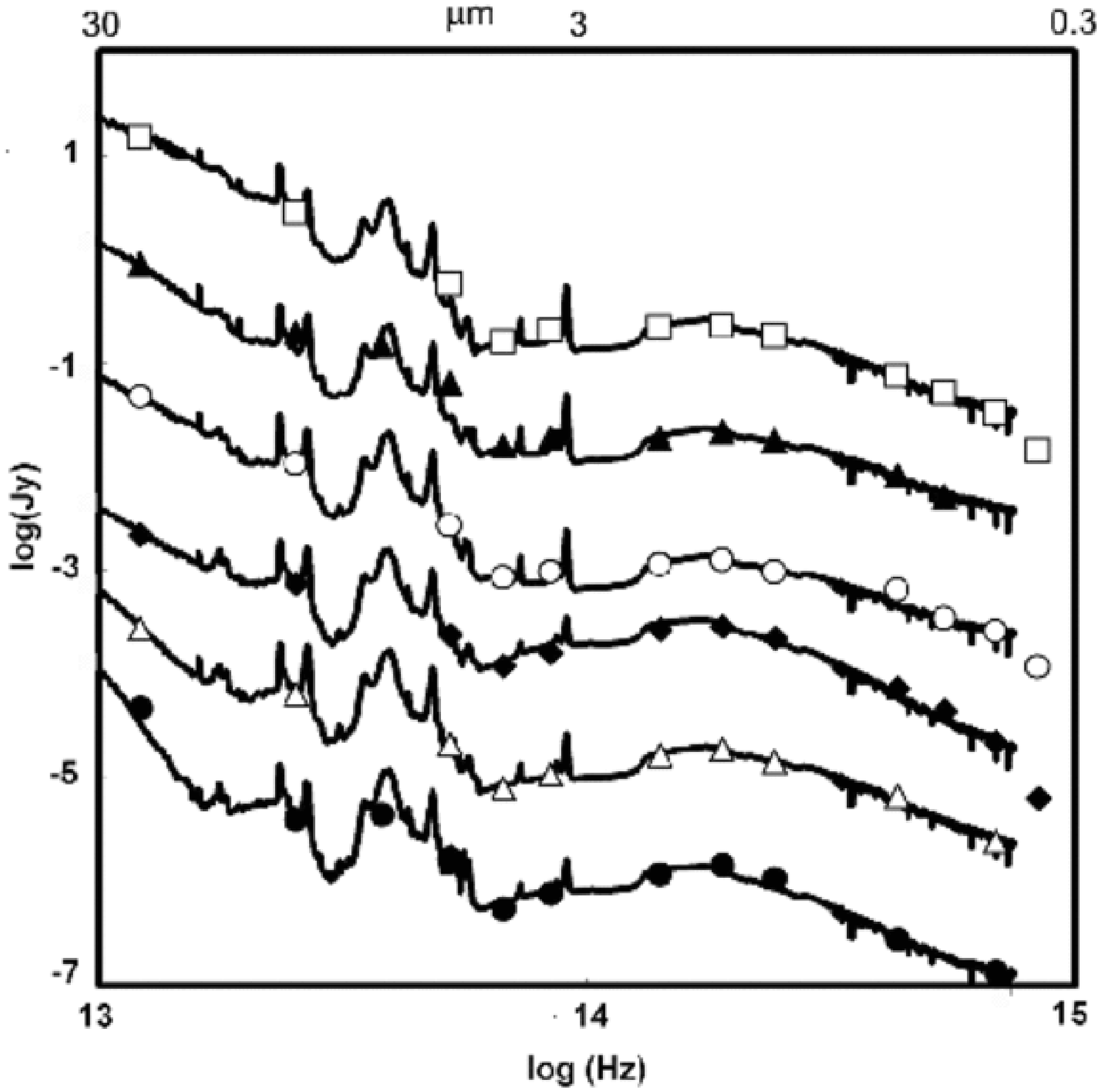}
\caption{Expanded LIRG SEDs to highlight optical through mid-IR. 
The order of galaxies and offsetting factors are as in Figure 1.  
\label{02}}
\end{figure}

\clearpage

\begin{figure}
\epsscale{1.0}
%\plotone{templates3}
\plotone{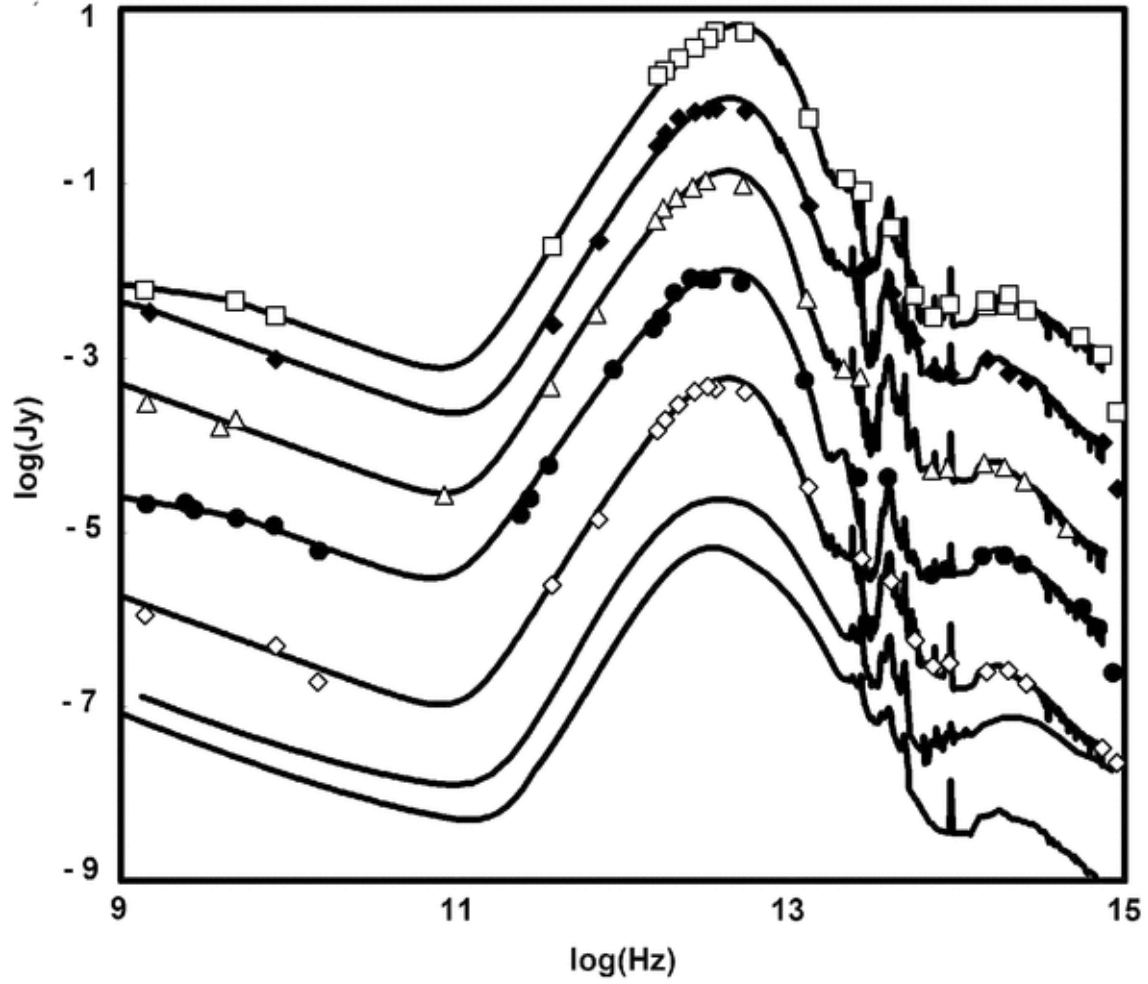}
\caption{Full ULIRG SEDs and defining photometric points. Between 5 and 36$\mu$m the SEDs are based on IRS spectra rather than the photometry. The curves have been offset for clarity; from top to bottom the galaxies and offset factors are: 1.) IRAS 22491-1808 (1); 2.) IRAS 14348-1447 (0.1); 3.) 
IRAS 17208-0018 (0.003); 4.) Arp 220 (0.00007); 5.) IRAS 12112+0305 (0.00005); 6.) Dale alpha = 1.5 model; and 7.) Chary \& Elbaz L(TIR) = $2 \times 10^{12}$ model.   
\label{03}}
\end{figure}

\clearpage

\begin{figure}
\epsscale{1.0}
%\plotone{templates4}
\plotone{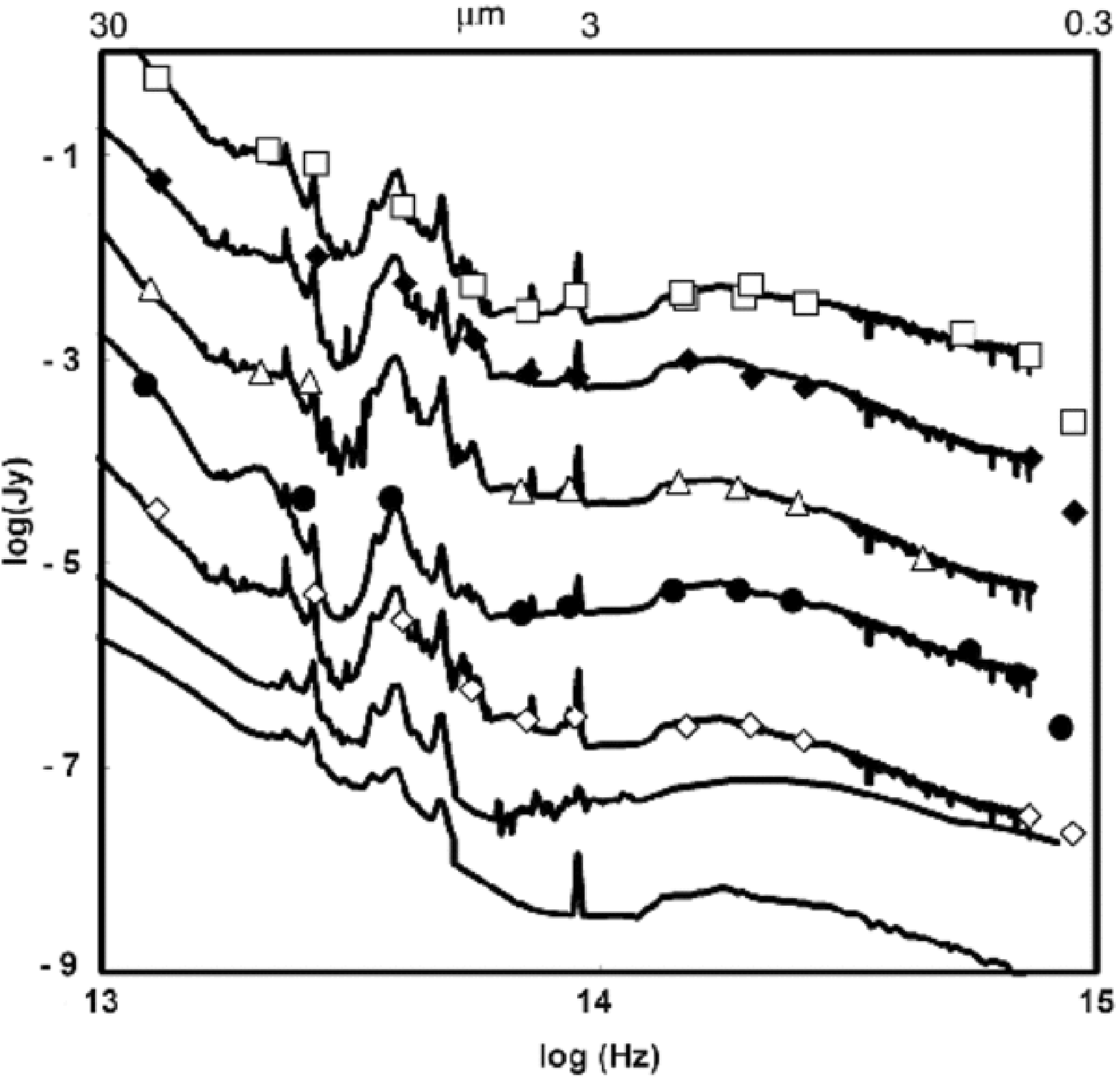}
\caption{Expanded ULIRG SEDs to highlight optical through mid-IR. 
The order of galaxies and offsetting factors are as in Figure 3. 
\label{04}}
\end{figure}

\clearpage

\begin{figure}
\epsscale{1.0}
%\plotone{template12}
\plotone{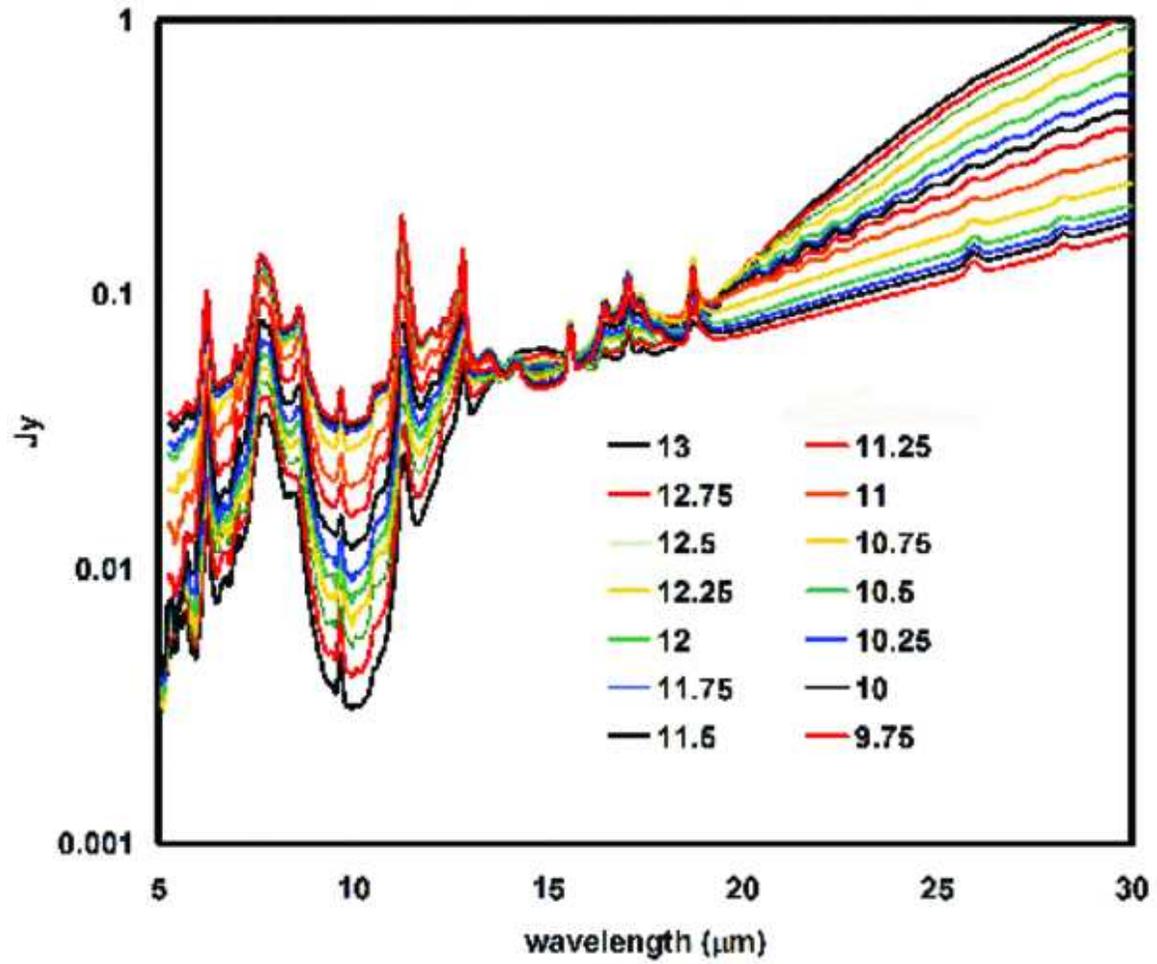}
\caption{Family of average spectral templates in the 5 to 37$\mu$m
range. These templates are used for K-correction to 24$\mu$m rest
band. The templates are keyed to the log(L(TIR)) of the galaxies, calculated
as by Sanders et al. (2003). They are normalized at 14$\mu$m.
The templates for the three highest luminosities are extrapolated 
and are likely to be significantly less representative than the rest.
\label{5}}
\end{figure}

\clearpage

\begin{figure}
\epsscale{1.0}
%\plotone{template18}
\plotone{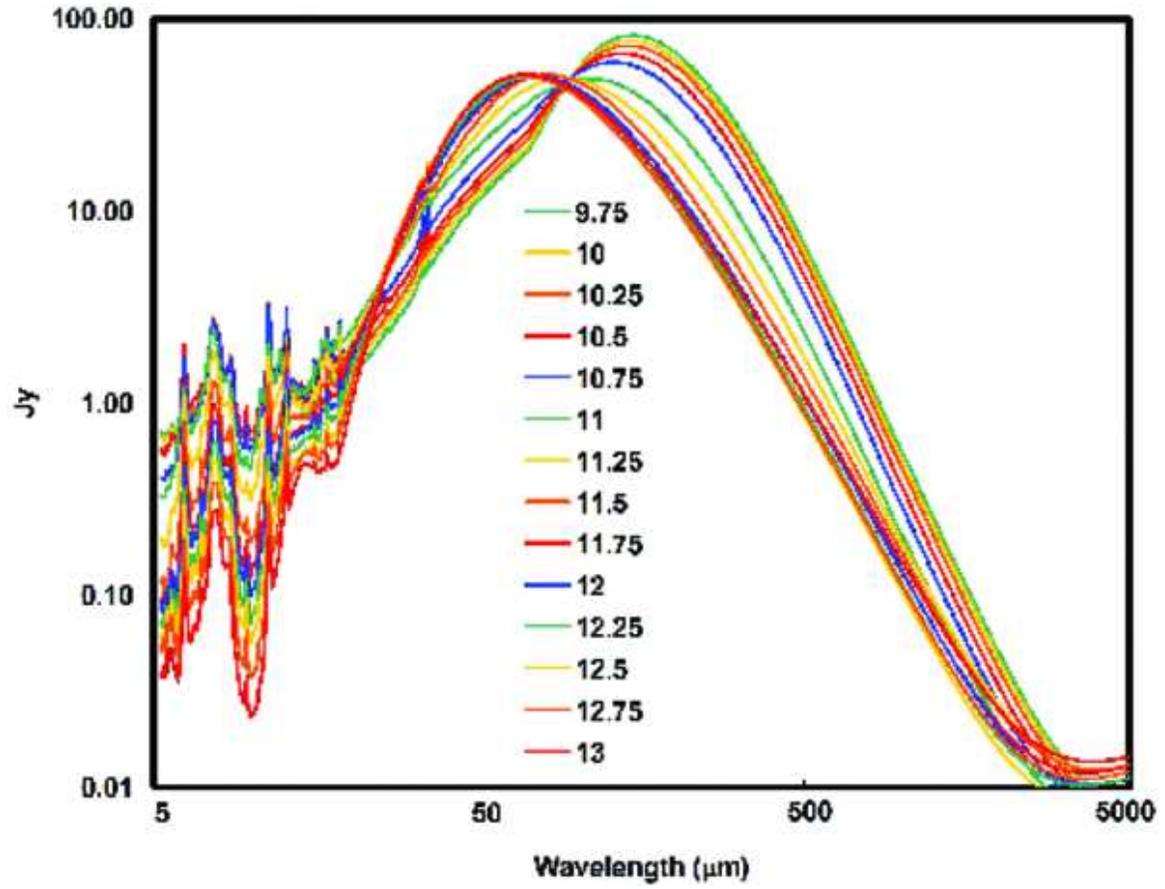}
\caption{Family of average templates for full radio, far-infrared, and mid-infrared
spectral range. The templates are keyed to the log(L(TIR)) of the galaxies, calculated
as by Sanders et al. (2003). They are normalized to the same integrated flux. 
The templates for the three highest luminosities are extrapolated and are 
likely to be significantly less representative than the rest.
\label{6}}
\end{figure}

\clearpage

\begin{figure}
\epsscale{1.0}
%\plotone{template17}
\plotone{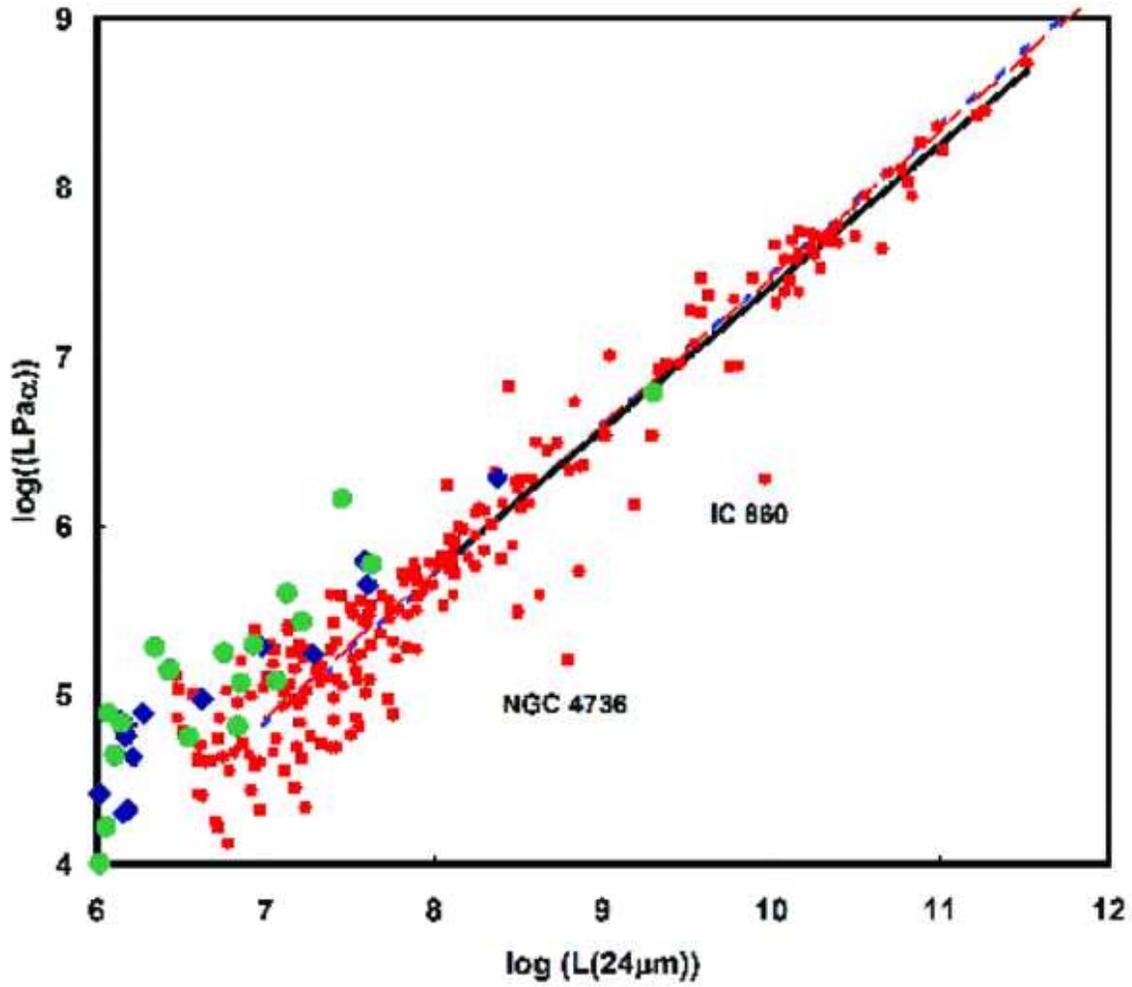}
\caption{Comparison of Pa$\alpha$ and 24$\mu$m luminosities. The red squares
are for high metallicity galaxies and HII regions as defined by Calzetti et al. (2007);
the green circles are for intermediate metallicity; and the blue diamonds are
for low metallicity. The fit (black line) is to the high metallicity points. The 
fits to subsets of the data by Alonso-Herrero et al. (2006) and Calzetti et al. (2007) 
are shown as dashed red and dashed blue lines, respectively.
\label{7}}
\end{figure}

\clearpage

\begin{figure}
\epsscale{1.0}
%\plotone{template21}
\plotone{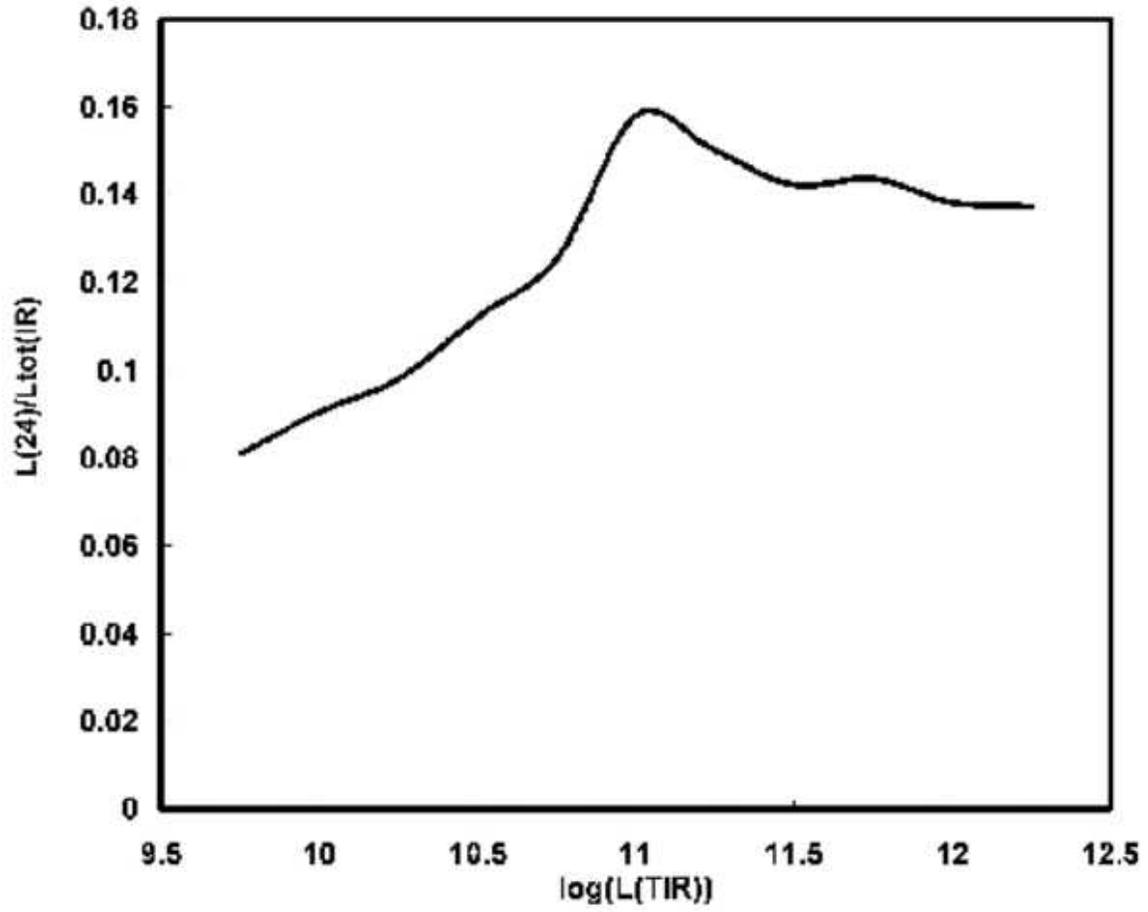}
\caption{Trend of L(24)/L$_{tot}$(IR) with L(TIR). L(TIR) is based
strictly on the formulation of Sanders et al. (2003) using IRAS data, while
L$_{tot}$(IR) is obtained by integrating under the infrared SED from
5 to 1000$\mu$m. 
\label{8}}
\end{figure}

\clearpage

\begin{figure}
\epsscale{1.0}
%\plotone{templates18}
%\plotone{z.lum24.bysfr.ps}
\plotone{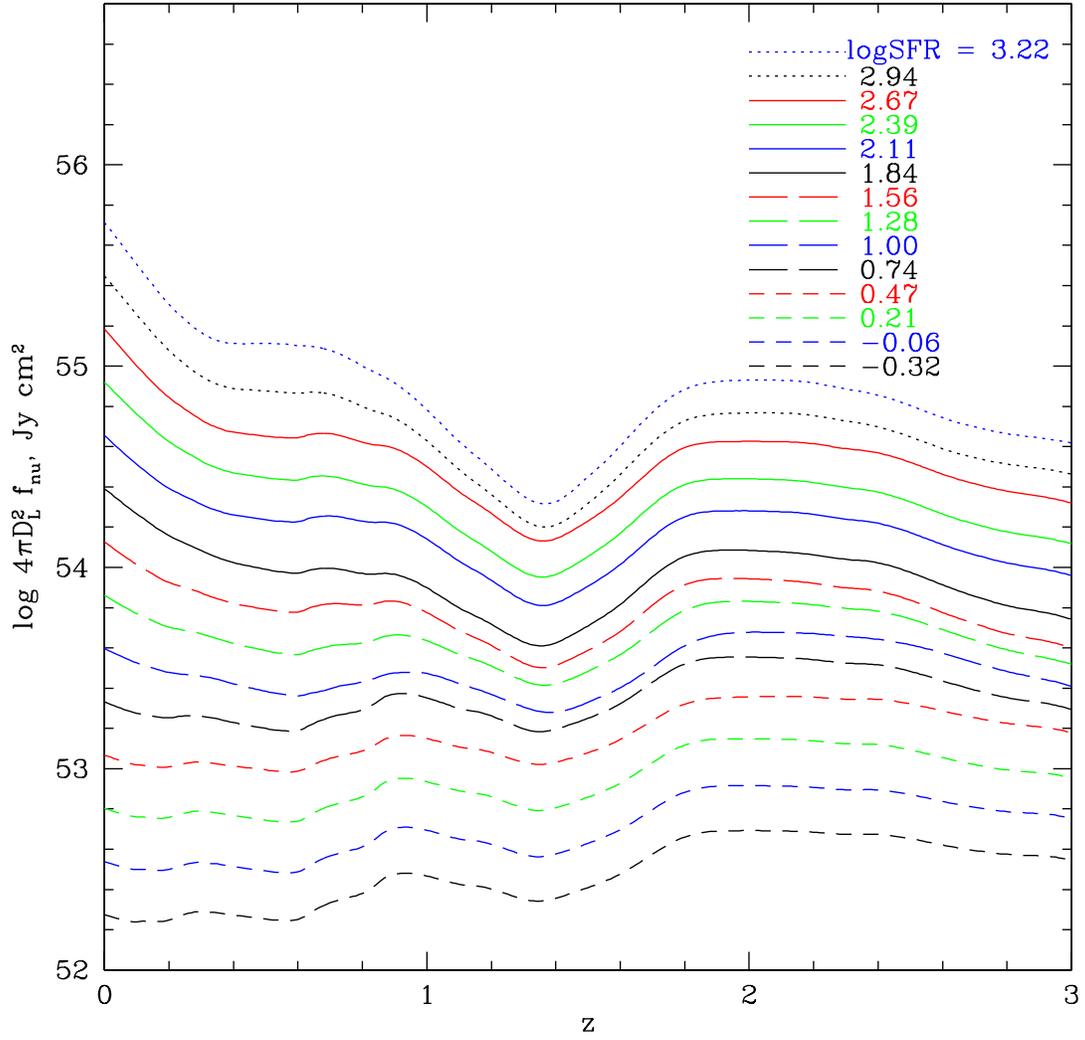}
\caption{Tracks of the average template SEDs vs. redshift, as would be 
observed at 24$\mu$m. The even spacing at any
given redshift implies that a power-law fit is a good approximation to the 
dependence of SFR on 24$\mu$m flux.
\label{9}}
\end{figure}

\clearpage

\begin{figure}
\epsscale{1.0}
%\plotone{templates19}
%\plotone{z.sfr_allfilt.fitpars.ps}
\plotone{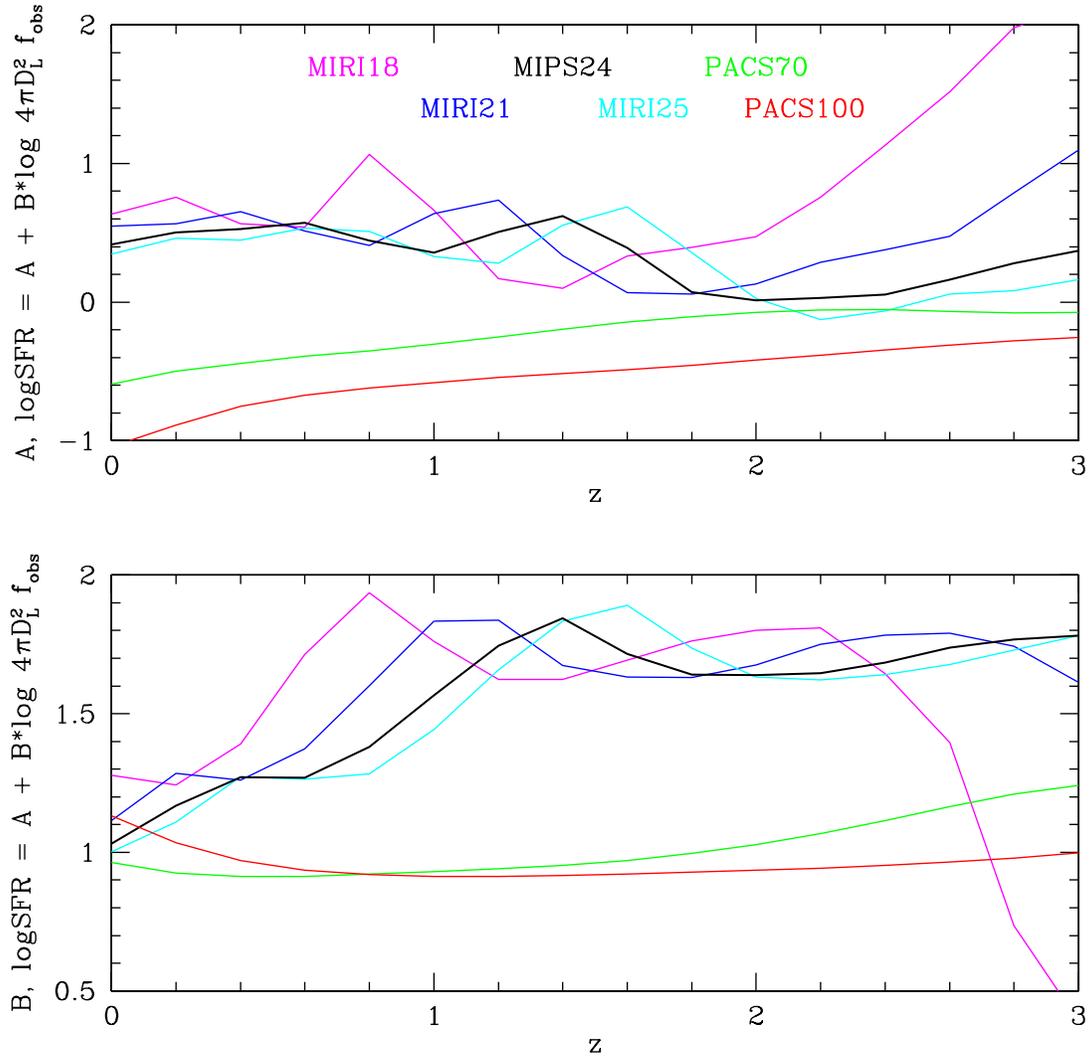}
\caption{Trends with redshift of the intercept $A$ and slope $B$ for the relation 
between SFR and infrared flux, ${\rm log~SFR} = A + B({\rm log}~4\pi D_L^2 f_\nu\, -\, 53)$, 
for several infrared bands: Spitzer/MIPS 24$\mu$m (heavy black line); Herschel/PACS
70 and 100$\mu$m (green and red lines); JWST/MIRI 18, 21, and 25$\mu$m (magenta, blue,
and cyan lines). 
\label{10}}
\end{figure}

\clearpage

\begin{figure}
\epsscale{1.0}
%\plotone{template22}
\plotone{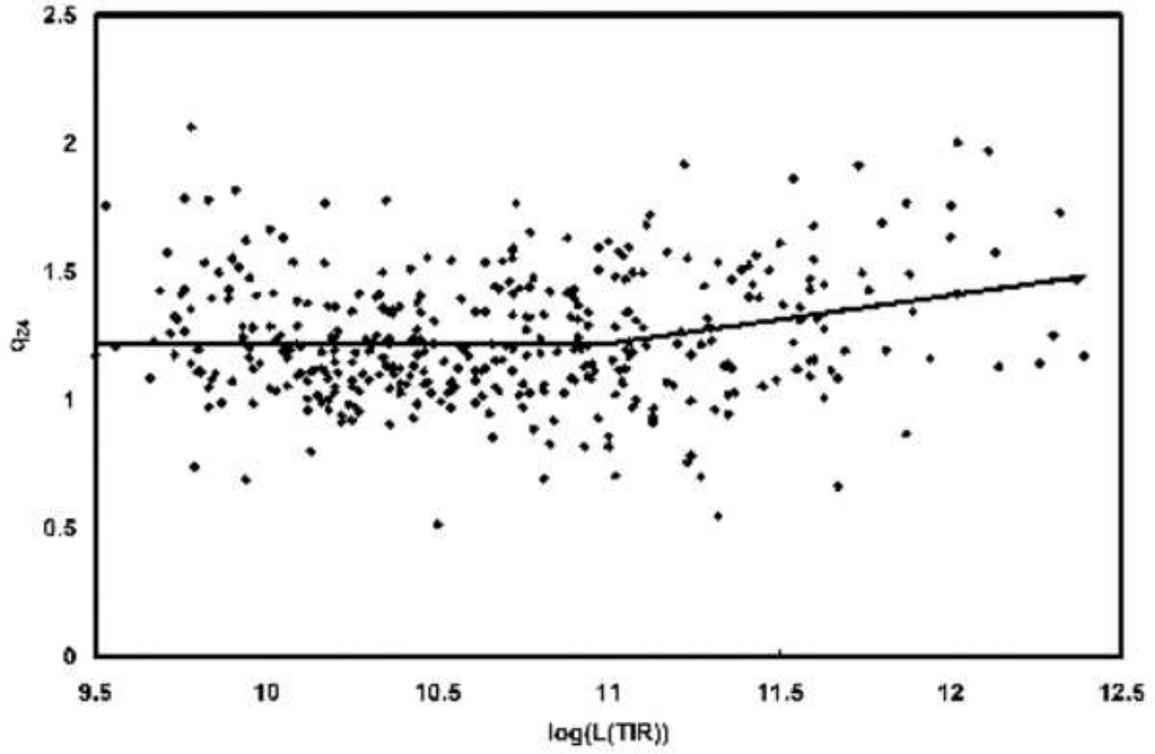}
\caption{Trend of q$_{24}$ = log[f$_\nu$(24$\mu$m)/f$\nu$(1.4GHz)] with log(L(TIR)).
The line shows our preferred fit, a constant of q$_{24}$=1.22 for log(L(TIR)) $<$ 11, and
a sloping line above. 
\label{A5}}
\end{figure}

\clearpage

\begin{figure}
\epsscale{1.0}
%\plotone{templates11}
\plotone{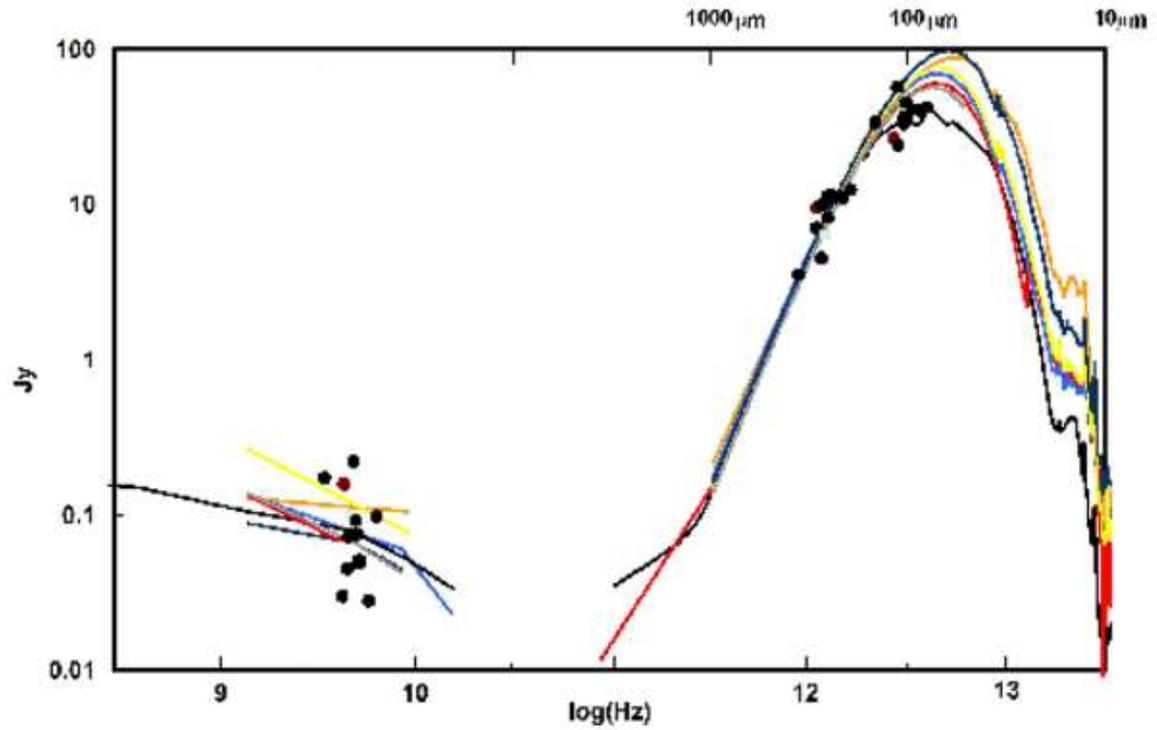}
\caption{Comparison of our templates with the
measurements of high redshift galaxies by Kov\'acs et al. (2006). 
The templates and the measurements
have all been normalized near 260$\mu$m, close to the rest wavelength
for the 850$\mu$m measurements of Kov\'acs et al. (2006). There is
substantial scatter in the resulting radio fluxes for the templates
but even more for the radio fluxes of the high redshift galaxies.
The latter are perhaps increased significantly be measurement errors
(not all of which are captured in the nominal errors, as demonstrated
by the results of independent reductions (Kov\'acs et al. 2006)). Nonetheless,
the local templates are very representative of the behavior of those
at high redshift.
\label{A8}}
\end{figure}

\clearpage

\begin{figure}
\epsscale{1.0}
%\plotone{template13}
\plotone{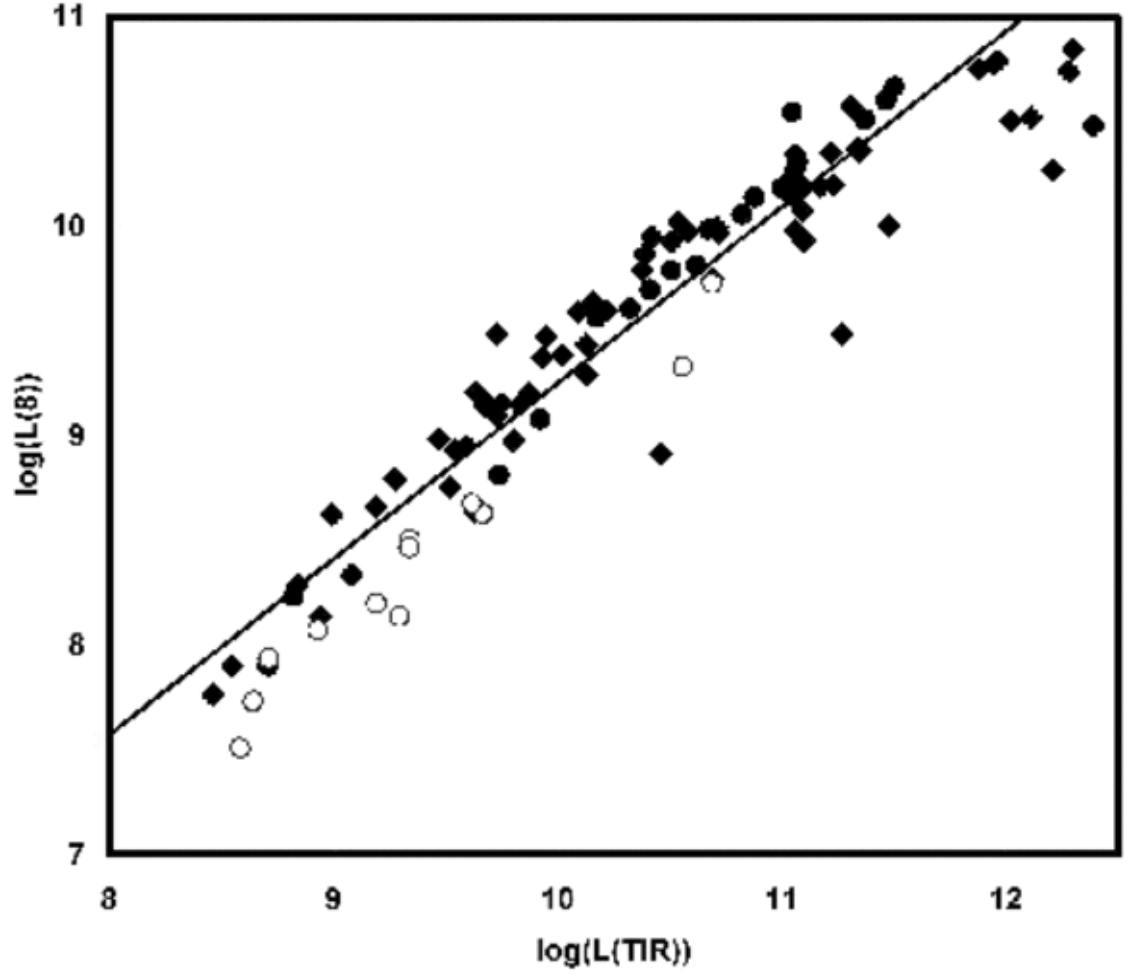}
\caption{Comparison of L(8) and L(TIR). Above log(L(TIR) = 11, the relation
appears almost to saturate due to a number of very luminous galaxies
with suppressed emission at 8$\mu$m. The diamonds are from Dale et al. (2007) and (for the LIRGs and ULIRGs) from our reduction of archival data. 
The circles are from Engelbracht et al. (2008); filled ones are within 0.25 dex of solar metallicity (in O/H) while open ones indicate metallicity between 0.25 and 0.5 dex below solar. The line is from Equation 21.
\label{A1}}
\end{figure}

\clearpage

\begin{figure}
\epsscale{1.0}
%\plotone{template15}
\plotone{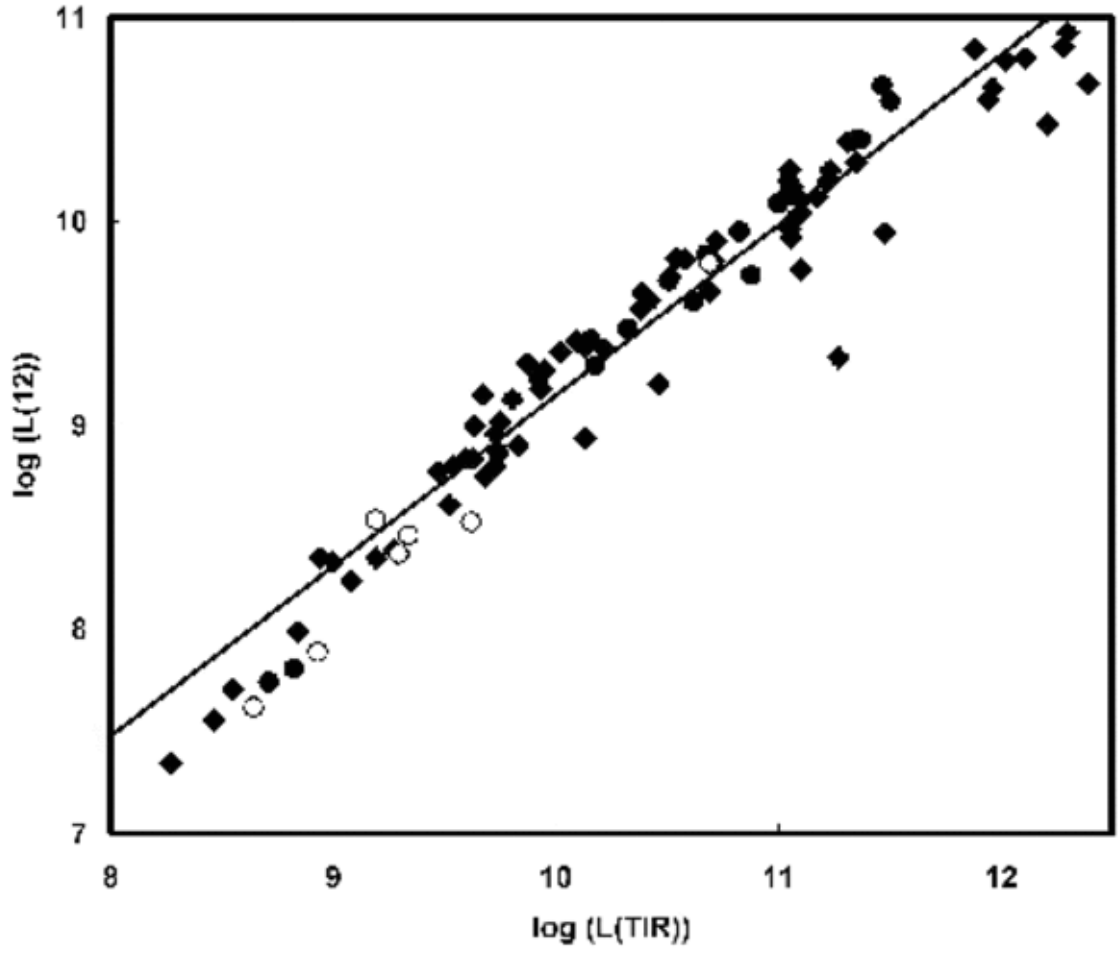}
\caption{Comparison of L(12) and L(TIR). The most deviant low point (NGC 1316) 
has been excluded from the fit. The slope is significantly different
from one (see Equation 23) and there are indications of an 
overall curvature, with a lower ratio of L(12)/L(TIR) at high luminosity. 
Symbols are as for Figure 13.
\label{A2}}
\end{figure}

\clearpage

\begin{figure}
\epsscale{1.0}
%\plotone{template14}
\plotone{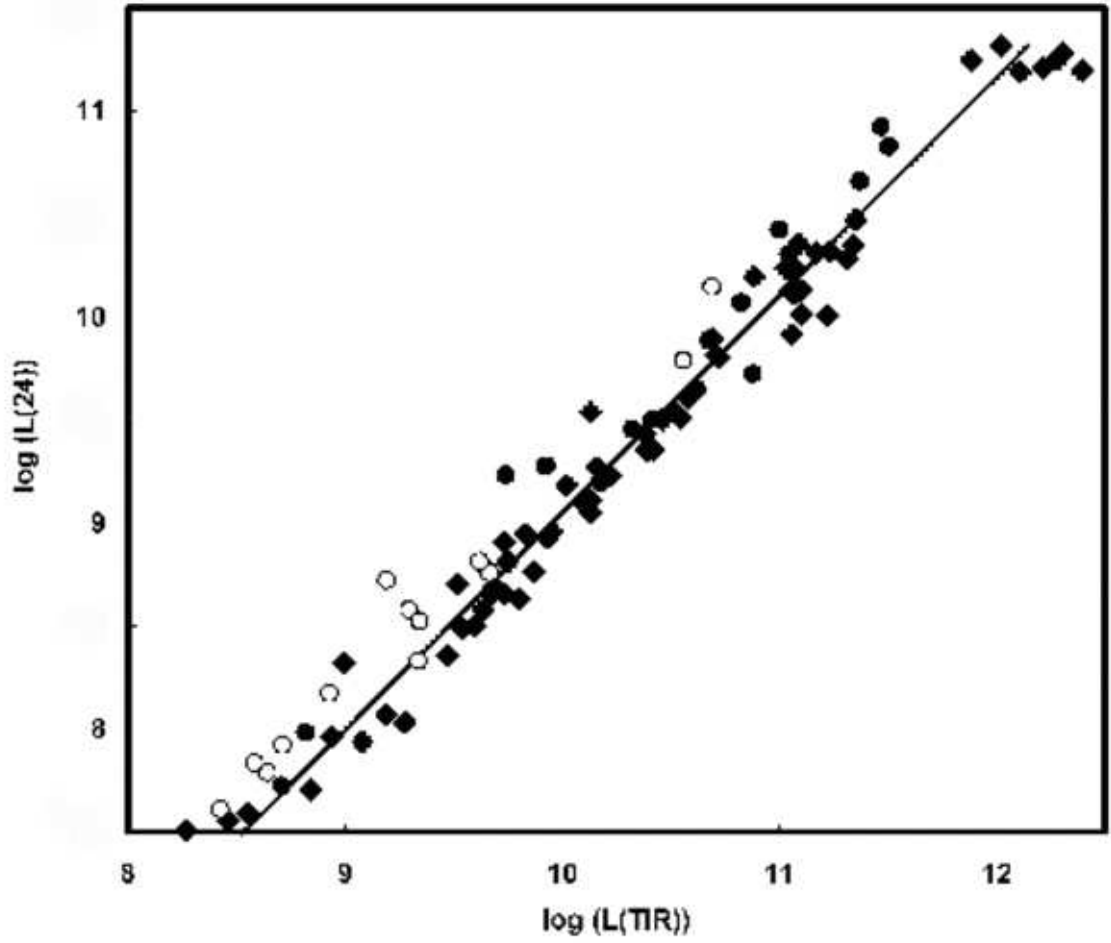}
\caption{Comparison of L(24) and L(TIR). The line is a linear fit
as given in the text. It fits well with virtually no outliers and
with a slope close to one (see Equation 25).  
Symbols are as for Figure 13. 
\label{A3}}
\end{figure}

\clearpage

\begin{figure}
\epsscale{1.0}
%\plotone{templates8}
\plotone{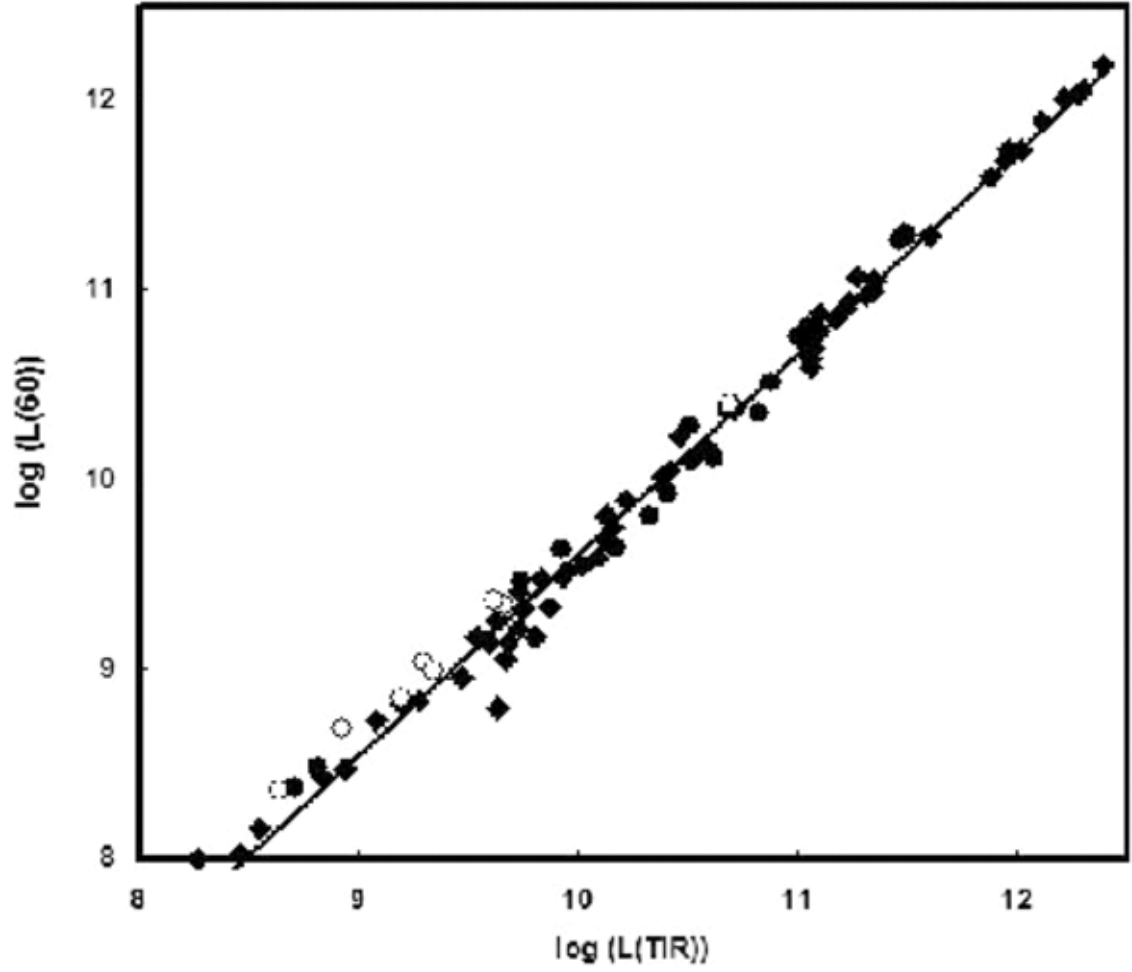}
\caption{Comparison of L(60) and L(TIR). The line is a linear
fit given in the text. There are no outliers (not surprisingly
because 60$\mu$m is close to the maximum of the SED) but
the slope differs somewhat from unity (see Equation 26).
Symbols are as for Figure 13.
\label{A4}}
\end{figure}

\clearpage

\begin{figure}
\epsscale{1.0}
%\plotone{template16}
%\plotone{fidel.z_f70f24.labels.tir5e9.ps}
\plotone{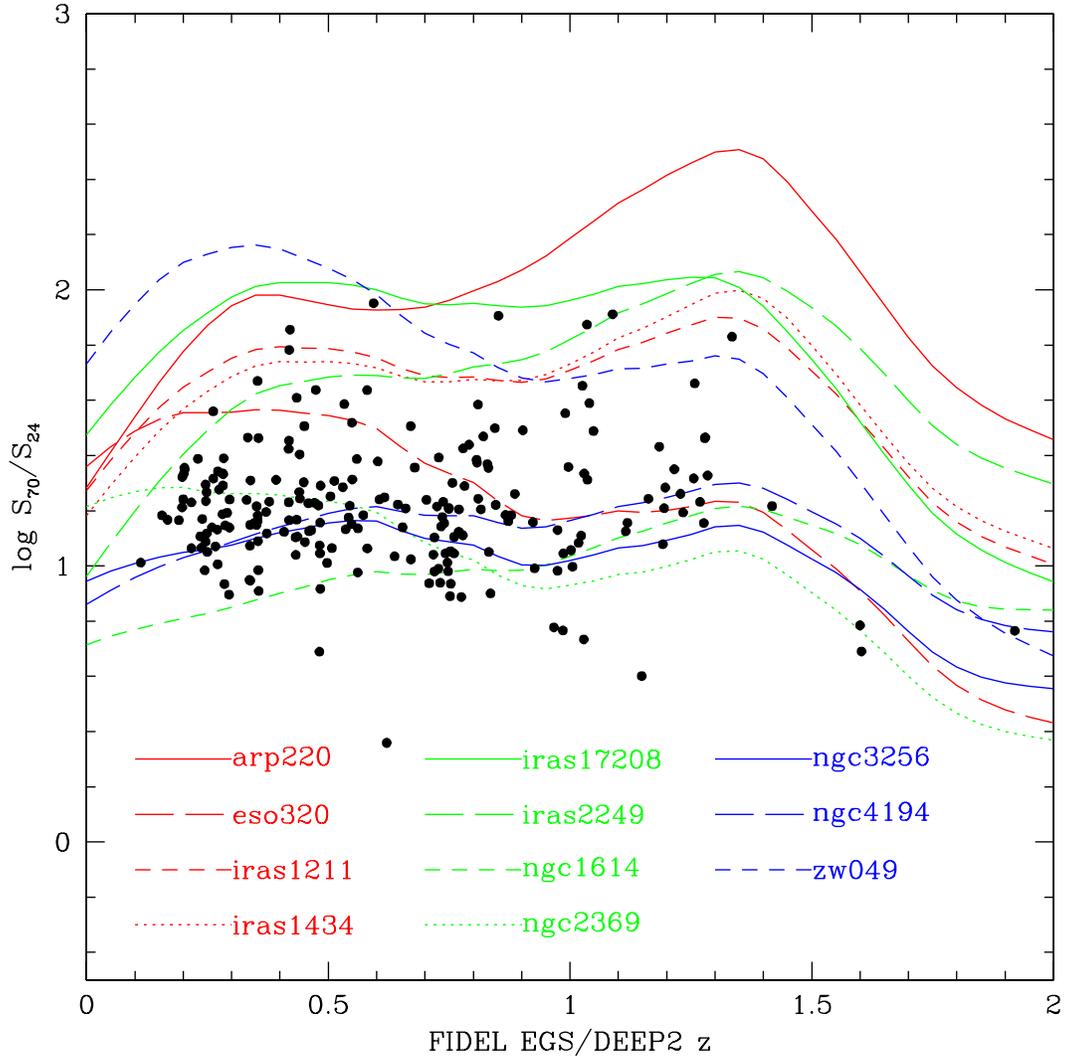}
\caption{Behavior of templates in the far infrared. The templates have
been redshifted and convolved with the MIPS relative response functions
to predict their behavior at 24 and 70$\mu$m. The points show the ratios
of 70 to 24$\mu$m flux densities in the EGS/FIDEL field for galaxies
with $L(TIR) > 5 \times 10^9$, and DEEP2 spectroscopic redshifts.  
Objects with low $S_{70}/S_{24}$ are mostly AGN. 
For $0 < z < 1$, the templates define the
space occupied by the measurements well. The "extra" template at low $z$ is Zw 49.057, 
which evidently represents an anomalous spectral energy distribution not
seen with any significant frequency in the FIDEL data. Above z $\sim$ 1,
the sensitivity limitations at 70$\mu$m should result in a substantial bias
in favor of sources with large 70/24 flux ratios and against those with
small ratios. Nonetheless, compared with the locus of the family of templates,
few such galaxies are identified. This result supports the work of
Papovich et al. (2007) and Rigby et al. (2008) that indicate an
increasing relative contribution of the $\sim$ 8$\mu$m region to the total
infrared outputs of galaxies at redshifts of z $\sim$ 2. 
\label{A6}}
\end{figure}

\clearpage

\begin{figure}
\epsscale{1.0}
%\plotone{template20a}
\plotone{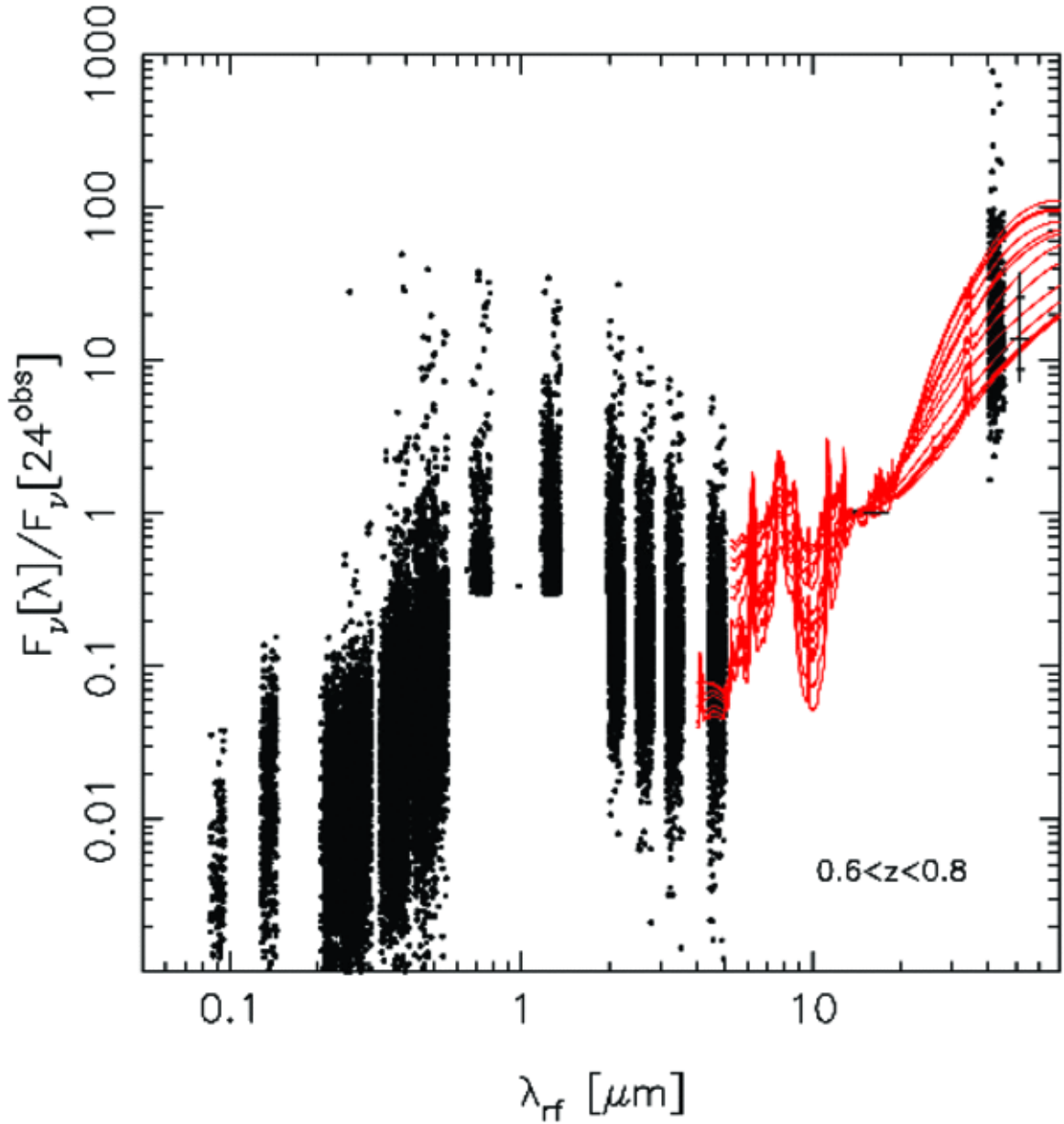}
\caption{Comparison of templates with photometry at z $\sim$ 0.7. 
The photometry is for galaxies in the EGS with [3.6] $<$ 23 (a
cut that is responsible for the truncation of the distribution of points
near 1~$\mu$m toward low values), photo-z between 0.6 and 0.8
(Perez-Gonzalez et al. 2008ab), a detection at 24~$\mu$m and/or 70~$\mu$m
(as measured in the DR2 FIDEL data; Dickinson et al. 2007), and an estimated
value of L(TIR) larger than 10$^{11}$~L$_\sun$. The vertical line and marks
on the right of the 70~$\mu$m fluxes show the median, quartiles and values
enclosing 68\% of the distribution of flux points. The templates
and photometry are all normalized at observed 24$\mu$m. 
\label{A7}}
\end{figure}

%\begin{figure}
%\epsscale{1.0}
%\plotone{templates9a}
%\caption{Behavior of Templates on the Stern Color-Color Diagram. 
%This figure is an obvious companion to the preceding one.
%\label{10}}
%\end{figure}

%\clearpage

%\clearpage

\end{document}